\documentclass[journal]{IEEEtran}

\usepackage{ifpdf}

\usepackage{cite}

\ifCLASSINFOpdf
  \usepackage[pdftex]{graphicx}  
\else
  \usepackage[dvips]{graphicx}
\fi
\graphicspath{{ }}

\usepackage{amsmath}     
\usepackage{enumitem}    
\usepackage{adjustbox}   
\usepackage{xcolor}      
\usepackage{hyperref}    
\usepackage{placeins}    
\usepackage{subcaption}  
\usepackage{float}       
\usepackage{caption}     
\usepackage{dblfloatfix} 
\usepackage{booktabs}

\usepackage{amsmath}

\usepackage{url}

\begin{document}
\hyphenation{op-tical net-works semi-conduc-tor}

\title{Array-Aware Ambisonics and HRTF Encoding for Binaural Reproduction With Wearable Arrays}

\author{Yhonatan~Gayer,~\IEEEmembership{Student~Member,~IEEE,} 
        Vladimir~Tourbabin, 
        Zamir~Ben-Hur, 
        David~Lou-Alon, and~Boaz~Rafaely,~\IEEEmembership{Fellow,~IEEE}%
\thanks{Yhonatan Gayer and Boaz Rafaely are with the School of Electrical and Computer Engineering, Ben-Gurion University of the Negev, Beer-Sheva, Israel.}
\thanks{Vladimir Tourbabin, Zamir Ben-Hur, and David Lou-Alon are with Reality Labs Research at Meta, Redmond, WA, USA.}%
}

%
%

\markboth{Journal of \LaTeX\ Class Files,~Vol.~14, No.~8, August~2015}%
{Shell \MakeLowercase{\textit{et al.}}: Bare Demo of IEEEtran.cls for IEEE Journals}
%

\maketitle

\begin{abstract}
This work introduces a novel method for binaural reproduction from arbitrary microphone arrays, based on array-aware optimization of Ambisonics encoding through Head-Related Transfer Function (HRTF) pre-processing. The proposed approach integrates array-specific information into the HRTF processing pipeline, leading to improved spatial accuracy in binaural rendering. Objective evaluations demonstrate superior performance under simulated wearable-array and head rotations compared to conventional Ambisonics encoding method. A listening experiment further confirms that the method achieves significantly higher perceptual ratings in both timbre and spatial quality. Fully compatible with standard Ambisonics, the proposed method offers a practical solution for spatial audio rendering in applications such as virtual reality, augmented reality, and wearable audio capture.

\end{abstract}

\begin{IEEEkeywords}
Ambisonics, Head-Related-Transfer-Function, Magnitude-Least-Squares, Binaural Reproduction.
\end{IEEEkeywords}

%
\IEEEpeerreviewmaketitle

\section{Introduction}
%
%
%
%
\IEEEPARstart{W}{ith} the rapid advancement and increasing adoption of virtual reality (VR) and augmented reality (AR) technologies, the demand for capturing acoustic environments and reproducing realistic spatial audio has grown significantly \cite{Review-Paper}. Spatial audio plays a crucial role in enhancing the immersive experience of VR and AR, providing users with a sense of directionality and depth that complements the visual elements.

Ambisonics \cite{book-Ambisonics} has emerged as a widely used format for spatial audio due to its flexibility and adaptability. By incorporating individualized Head-Related Transfer Functions (HRTFs) \cite{HRTF_measurements}, Ambisonics enables precise binaural reproduction over headphones, tailored to the listener's unique auditory characteristics. Furthermore, it seamlessly accommodates compensation for the listener head movements by applying rotations using the Wigner-D Matrix \cite{Wigner-D}, making it particularly suitable for dynamic VR and AR applications.

Traditionally, Ambisonics signals are computed using spherical microphone array recordings through the Plane Wave Decomposition (PWD) technique \cite{PWD-with-Spherical-conv}. However, this method necessitates the use of specially designed microphone arrays \cite{SH_Processing-book}, \cite{em32}, typically of a spherical configuration, which may not be practical for mobile or wearable devices commonly used in VR and AR settings \cite{Spherical-Ambisonics}. These limitations highlight the need for more versatile approaches to Ambisonics encoding that can accommodate arbitrary microphone array configurations.

To address these challenges, prior works \cite{Vlad-paper, Parametric-ASM-like-paper} have proposed methods for encoding Ambisonics signals from arbitrary array geometries. The method described in \cite{Parametric-ASM-like-paper} employs a parametric audio encoding stage, while \cite{Vlad-paper} adopts a simpler approach based on solving linear equations for Ambisonics coefficients. While these methods show promise, they also face limitations, such as errors introduced by dependencies on scene-specific estimated parameters, that may be difficult to estimate accurately, in particular with complex and dynamic acoustic scenes. Furthermore, conventional linear encoding as in \cite{Vlad-paper} may also have limitations due to the microphone array configuration which may not support accurate encoding of all Ambisonics channels \cite{DOA-non-symmetric-shape}.
The recent work in \cite{ASM} proposed improvement over the conventional linear encoding, but this came at the expense of additional channels.

Alternative approaches outside the Ambisonics framework have explored direct optimization of binaural signals using the Magnitude Least Squares (MagLS) \cite{book-Ambisonics} criterion. One method \cite{endToEndMagLS} applies this technique to spherical and circular microphone arrays, while another \cite{BSM_journal_paper} extends it to arbitrary array geometries, including wearable configurations. Although both methods achieve accurate reproduction, they do not perform Ambisonics encoding and are not compatible with standard Ambisonics pipelines.

This paper focuses on encoding Ambisonics signals from arbitrary microphone array configurations using a signal-independent approach. By avoiding reliance on the acoustic scene, this method eliminates the need to estimate signal or scene-specific parameters, and may potentially be suitable for challenging acoustic scenes.

The proposed method is based on the observation that the encoding process may lead to two types of errors: (1) truncation error, resulting from the limited number of encoded Ambisonics channels, and (2) intrinsic encoding errors within the estimated channels. These limitations can substantially degrade the quality of binaural signals decoded from Ambisonics, highlighting the need for a more robust encoding-decoding process that can accommodate such imperfections.

To address truncation-related degradation, prior works \cite{HRTF_MagLS}, \cite{berebi2023imagls} proposed HRTF preprocessing methods that improve binaural reproduction even with a small number of Ambisonics channels, while \cite{Engel2022} and \cite{Ambisoncis_and_BSM_comparison} provided evaluations of these approaches. However, these solutions primarily target truncation errors, without accounting for inaccuracies due to the encoding process, stemming from the array geometry or the specific Ambisonics encoding used.

This paper first analyzes the limitations of existing Ambisonics encoding approaches when applied to a wearable array. Then, it introduces a novel HRTF preprocessing framework that explicitly accounts for both the Ambisonics truncation error, and error stemming from the encoding process due to the geometry of the array, thereby improving the accuracy and perceptual quality of binaural reproduction. \color{black}

The key contributions of this paper are outlined as follows:
\begin{itemize}[left=0em, topsep=0pt, itemsep=-0.3ex]
\item A novel theoretical and numerical analysis, highlighting the limitations of Ambisonics encoding with wearable arrays and its implications for binaural reproduction. 
\item Development and investigation of a novel target error function that explicitly incorporates microphone array geometry and encoder properties. This function is designed to minimize HRTF-related reproduction errors and enhance the perceptual quality of binaural signals derived from encoded Ambisonics.
\item Demonstration of the proposed approach using a wearable microphone array configuration. \color{black}
\item Analysis of performance of the proposed method through numerical simulations and a listening test.
\end{itemize}

\section{Background}

This section establishes the mathematical basis for the signal model, introduces the Ambisonics representation based on this model, and describes the methods for Ambisonics encoding and binaural reproduction from both spherical and arbitrary microphone arrays. 

\subsection{Signal Model}
Consider an arbitrary array comprising $M$ omni-directional microphones, each positioned at coordinates $(r_i, \theta_i, \phi_i)$, $\forall 1 \le i\le M$. Consider also a set of $Q$ plane waves with directions of arrival (DOA) $(\theta_q, \phi_q) \hspace{2mm} \forall \hspace{2mm} 1 \le q \le Q$, denoted as $\Omega_Q$. The array steering matrix is denoted as $\mathbf{V}(k)$ with dimensions $M \times Q$, where each element $\left[ \mathbf{V}(k) \right]_{i,q}$ corresponds to the frequency response of the $i$-th microphone to a plane wave arriving from the DOA $(\theta_q,\phi_q)$ at wave number $k$.
The signal measured by the microphones can be expressed as:
\begin{equation}
    \textbf{x}(k) = \mathbf{V}(k)\mathbf{s}(k) + \mathbf{n}(k)
\label{eq:mic model}
\end{equation}
where $\mathbf{x}(k) = [ x_1(k) , ... , x_M(k) ]^T$ is a vector of length $M$, each element $x_i(k)  \hspace{2mm} \forall 1 \le i \le M$ represents the signal captured by the $i$-th microphone. $\mathbf{s}(k) = [ s_1(k) ,\dots, s_Q(k) ]^T$ is the sources signal vector of size $Q$, where each element represents the amplitude of a plane wave at the origin. Finally, $\mathbf{n}(k) = [ n_1(k) , ... , n_M(k) ]^T$ is the microphone noise vector of size $M$, assumed to be independently, identically distributed (i.i.d.) and uncorrelated with $\mathbf{s}(k)$.

Additionally, the binaural signal at the left or right ear can be represented utilizing the space domain representation of the HRTF, denoted by $\mathbf{\textbf{h}}^{l,r}(k) = [h^{l,r}(\theta_1,\phi_1,k), \dots, h^{l,r}(\theta_Q,\phi_Q,k)]^T$ of size $Q$, and the source vector $\textbf{s}(k)$ as in (\ref{eq:mic model}):
\begin{equation}
    p^{l,r}(k) = \left[\mathbf{h}^{l,r}(k)\right]^T\mathbf{s}(k)
    \label{eq:p=hs}
\end{equation}
\subsection{Ambisonics}
\label{section:Ambisonics}
The Ambisonics signal due to $\mathbf{s}(k)$ in (\ref{eq:mic model}) and the $Q$ plane waves can be represented as follows, see (2.43) in \cite{SH_Processing-book}:
\begin{equation}
\mathbf{a_{nm}}(k) = \mathbf{Y}_{\mathbf{\Omega}_Q}^H \mathbf{s}(k)
\label{eq:anm=Y^Hs}
\end{equation}
Here, $\mathbf{Y}_{\mathbf{\Omega}_Q} = [\mathbf{y_{00}},\dots,\mathbf{y_{N_aN_a}}]$ denotes the spherical harmonics (SH) matrix of size $Q \times (N_a+1)^2$, where $\mathbf{y_{nm}} = [Y_{nm}(\theta_1,\phi_1),...,Y_{nm}(\theta_Q,\phi_Q)]^T$, $\forall \hspace{2mm} 0 \leq n \leq N_a$, $-n \leq m \leq n$, denoting a vector of size $Q$ that holds the SH functions of order $n$ and degree $m$, at $(\theta_q,\phi_q)$, see Chapter 1 of \cite{SH_Processing-book}.
Additionally, $\mathbf{a_{nm}}(k) = [a_{00}(k), \ldots, a_{N_aN_a}(k)]^T$ has a size of $(N_a+1)^2$ and holds the Ambisonics signals up to order $N_a$.

Ambisonics signals can be employed for rendering binaural signals \cite{book-Ambisonics}, \cite{Ambisonics2binaurals}, \cite{hrft_Ambisonics}:
\begin{equation}
    p^{l,r}(k) = \left[ \mathbf{h}_\mathbf{nm}^{l,r} (k)\right]^T\tilde{\mathbf{a}}_{\mathbf{nm}}(k)
    \label{eq:p=hnmTanm}
\end{equation}
where \( p^{l,r}(k) \) denotes the binaural signal. In this context, 
\(\mathbf{h}_\mathbf{nm}^{l,r}(k) = [h_{00}(k), \dots, h_{N_hN_h}(k)]^T\) is a vector of size $(N_h+1)^2$ representing the HRTF in the SH domain, and $\tilde{\mathbf{a}}_{\mathbf{nm}}(k) = [\tilde{a}_{00}(k), ..., \tilde{a}_{N_aN_a}(k)]$.
It is important to note that \(\tilde{\mathbf{a}}_\mathbf{nm}^{l,r}(k)\) is a rearranged and sign-inverted version of the original Ambisonics vector, where \(\tilde{a}_{nm}(k) = (-1)^m a_{n,-m}(k)\).
For (\ref{eq:p=hnmTanm}) to be applicable, we truncate either $\Tilde{\mathbf{h}}_\mathbf{nm}^{l,r}(k)$ or $\mathbf{a_{nm}}(k)$ so that \( N_a = N_h \), as described in \cite{book-Ambisonics}, \cite{Vlad-paper}.

Ambisonics channels can be computed with high accuracy when utilizing spherical microphone arrays that are uniformly or nearly uniformly distributed and mounted on either a rigid or open sphere \cite{book-Ambisonics}. In such cases, Ambisonics encoding is performed using Plane Wave Decomposition (PWD), which ensures accurate computation of the Ambisonics coefficients  \(\forall n \leq N_a , \forall -n \leq m \leq n \), provided the following condition is met:
\begin{equation}
    (N_a + 1)^2 \leq M
    \label{cond:(Na+1)^2<M}
\end{equation}
This ensures a total of \( (N_a+1)^2 \) accurate Ambisonics channels in the operating frequency range of the spherical array \cite{rafaely2004analysis}.

\subsection{Compensation for Head Rotation}

In AR and VR applications, immersive audio relies on the listener’s ability to rotate their head while perceiving a stable, world-locked auditory scene. To achieve this ability, binaural signals must dynamically adapt to head movements. With Ambisonics-based binaural reproduction, i.e. Eq.(\ref{eq:p=hnmTanm}), this adaptation can be achieved by either rotating the HRTF to match the new head orientation, or by counter-rotating the Ambisonics signal.




By using the formulation of rotation of function on the sphere in the SH domain \cite{roation_of_SH}, and for simplicity assuming a 2D rotation of $\Delta\phi$ in the azimuth and $\Delta\theta$ in the elevation, Eq. (\ref{eq:p=hnmTanm}) can be rewritten using rotation expressed by the Wigner-D matrix \cite{SH_Processing-book}
\begin{equation}
\begin{split}
    p^{l,r}_{rot}(k) &= \left[ \mathbf{D}(\Delta\phi,\Delta\theta,0)\hspace{1mm}  \mathbf{h}_\mathbf{nm}^{l,r}(k)  \right]^T \tilde{\mathbf{a}}_{\mathbf{nm}}(k)\\
    &= \left[ \mathbf{h}_\mathbf{nm}^{l,r}(k)\right]^T  \mathbf{D}(\Delta\phi, \Delta\theta,0)^T \tilde{\mathbf{a}}_{\mathbf{nm}}(k)
    \label{eq:p=hnmDanm}
\end{split}
\end{equation}
where $p^{l,r}_{rot}(k)$ represents the binaural signals adapted to the listener’s head rotation, by either rotating the HRTF or the Ambisonics signal, and $ \mathbf{D}(\Delta\phi, \Delta\theta, 0)^T = \mathbf{D}(0, -\Delta\theta, -\Delta\phi)$ representing the counter rotation.

\subsection{HRTF Encoding using MagLS}
\label{sec:HRTFmagLS}
To improve the accuracy and perceptual fidelity of binaural reproduction using Ambisonics signals, effective preprocessing of HRTFs in the SH domain is essential. Among various techniques, Magnitude Least Squares (MagLS) optimization has proven particularly effective at high frequencies, where it helps mitigate perceptual artifacts arising from truncation errors associated with low-order Ambisonics.

The MagLS method, initially proposed in~\cite{HRTF_MagLS}, seeks to minimize the magnitude error between a reference HRTF and its SH-domain representation. Given the reference HRTF, $\mathbf{h}^{l,r}(k)$, the MagLS error is defined as:
\begin{equation}
\epsilon_{\text{bin}}^{\text{MagLS}} = \bigg\Vert \big|[\mathbf{h}_{\mathbf{nm}}^{l,r}(k)]^T\mathbf{Y}_{\Omega_Q}^T\big| - \big|[\mathbf{h}^{l,r}(k)]^T\big|\bigg\Vert^2 
\label{eq:epsilon_MagLS}
\end{equation}

Here, $\epsilon_{\text{bin}}^{\text{MagLS}}$ denotes the binaural magnitude error, and $|\cdot|$ represents the element-wise absolute value. The SH-domain coefficients that minimize this error are obtained by solving:
\begin{equation}
\mathbf{h}_{\mathbf{nm},\text{MagLS}}^{l,r}(k) = \arg\min_{\mathbf{h}_{\mathbf{nm}}^{l,r}(k)} \epsilon_{\text{bin}}^{\text{MagLS}} 
\label{eq:argmin_epsilon_MagLS}
\end{equation}

This approach is particularly suited to the high frequencies, where human auditory perception is dominated by magnitude.~\cite{berebi2023imagls}.

Finally, HRTF representation adopted in this work combines the conventional SH-domain representation at low frequencies, with its MagLS-optimized counterpart at high frequencies:
\begin{equation}
\tilde{\mathbf{h}}^{l,r}_{\mathbf{nm}, \text{final}}(k) = (1 - \alpha(k)) \tilde{\mathbf{h}}^{l,r}_{\mathbf{nm}}(k) + \alpha(k) \tilde{\mathbf{h}}^{l,r}_{\mathbf{nm}, \text{MagLS}}(k)
\label{eq:hnm_weighted}
\end{equation}

Here, $\alpha(k)$ is a frequency-dependent weighting function, gradually transitioning from 0 to 1 in the range $k_{\text{cutoff min}} < k < k_{\text{cutoff max}}$, defined as:
\begin{equation}
\alpha(k) =
\begin{cases} 
0, & k \leq k_{\text{cutoff min}} \\
\frac{k - k_{\text{cutoff min}}}{k_{\text{cutoff max}} - k_{\text{cutoff min}}}, & k_{\text{cutoff min}} < k < k_{\text{cutoff max}} \\
1, & k \geq k_{\text{cutoff max}}
\end{cases}
\label{eq:alpha}
\end{equation}

The selection of $\alpha(k)$, $k_{\text{cutoff min}}$, and $k_{\text{cutoff max}}$ is informed by perceptual considerations. In this work, a linear transition from 0 to 1 is applied between $f_{\min} = 800$\,Hz and $f_{\max} = 1.3\,$kHz. Note that $\alpha(k)$ may also be defined directly in terms of frequency in Hz.

MagLS preprocessing was shown to effectively mitigate truncation errors that impair localization and timbral accuracy when using low Ambisonics orders $N_a$. The underlying nonlinear optimization, based on magnitude-only fitting, is typically solved via iterative methods as elaborated in~\cite{kassakian2006convex}, with practical adaptations ensuring inter-frequency phase consistency~\cite{book-Ambisonics}.

\subsection{Ambisonics Encoding from Arbitrary Arrays}
\label{section: Ambisonics Encoding from Arbitrary Arrays}
To achieve a more generalized approach for binaural reproduction from arbitrary arrays, it is feasible to encode Ambisonics channels directly from the array.
Ambisonics are typically encoded from spherical arrays or specifically designed array \cite{book-Ambisonics}, \cite{SH_Processing-book}, \cite{em32}. However, for encoding Ambisonics using arbitrary array configurations as in \cite{Vlad-paper} and \cite{Parametric-ASM-like-paper}, we draw inspiration from binaural signal matching (BSM) \cite{BSM-conference-paper}, which utilizes Tikhonov regularization \cite{Tikhinov}. This approach is based on linear mapping from the microphone signals to the Ambisonics signal:
\begin{equation}
\begin{split}
    \hat{a}_{nm}(k) = \mathbf{c}_{nm}(k)^H \mathbf{x}(k), \\
    \forall \, 0 \leq n \leq N_a, \, -n \leq m \leq n
\label{eq:anm_hat=Casmx}
\end{split}
\end{equation}
where $\hat{\mathbf{a}}_{\mathbf{nm}}(k) = [\hat{a}_{00}(k),\dots,\hat{a}_{N_aN_a}(k)]^T$ denotes the estimated Ambisonics vector, $\mathbf{a_{nm}}(k) = [a_{00}(k), \dots, a_{N_aN_a}(k)]^T$ of length $(N_a+1)^2$.
This approach entails minimizing the following normalized mean squared error (NMSE) function to compute the optimal coefficients:
\begin{equation}
    \begin{aligned}
        \scalebox{1.5}{\(\varepsilon\)}_{\text{ASM}} = E\left[\left\lVert \hat{a}_{nm}(k) - a_{nm}(k)\right\rVert_2 ^2\right]\bigg/E\left[\left\lVert a_{nm}(k) \right\rVert_2 ^2\right]
    \end{aligned}
    \label{eq:error_nm}
\end{equation}
Aiming to minimize (\ref{eq:error_nm}) we substituting (\ref{eq:mic model}), (\ref{eq:anm=Y^Hs}) and (\ref{eq:anm_hat=Casmx}) 
into (\ref{eq:error_nm}): \color{black}
\begin{equation}
    \scalebox{1.5}{\(\varepsilon\)}_{\text{ASM}} = \frac{E\left[\left\lVert [\mathbf{c}_{nm}(k)]^H\big(\mathbf{V}(k)\mathbf{s}(k) + \mathbf{n}(k)\big) - [\mathbf{y}_{nm}]^H\mathbf{s}(k)\right\rVert_2 ^2\right]}{E\left[\left\lVert [\mathbf{y}_{nm}]^H\mathbf{s}(k) \right\rVert_2 ^2\right]}
\end{equation}
We assume that the microphone noise $\mathbf{n}(k)$ is white such that $\mathbf{R}_n = E[\mathbf{n}(k)\mathbf{n}(k)^H] = \sigma_n^2 \mathbf{I}$, and is uncorrelated with the source $\mathbf{s}(k)$, and that $\mathbf{R_s}(k) = \sigma_s^2\mathbf{I}$ which corresponds to qualities of a diffuse sound field composed of $Q$ plane waves arriving from DOAs $(\theta_1,\phi_1),...,(\theta_Q,\phi_Q)$ with equal magnitude, random phase, and uncorrelated with each other leading to:
\color{black} 
\begin{equation}
    \scalebox{1.5}{\(\varepsilon\)}_{\text{ASM}} = \frac{\sigma_s^2 \left\lVert\mathbf{V}(k)^H\mathbf{c_{nm}}(k) - \mathbf{y_{nm}}\right\rVert_2^2 + \sigma_n^2 \left\lVert\mathbf{c_{nm}}(k)\right\rVert_2^2}{\sigma_s^2\left\lVert\mathbf{y_{nm}}\right\rVert_2^2}
\label{eq:Ambisonics analytical error}
\end{equation}
Solving for the optimal encoder coefficients, that minimize the error expression in (\ref{eq:Ambisonics analytical error}) leads to the Ambisonics Signal Matching (ASM) solution:
\begin{equation}
    [\mathbf{c}_{\mathbf{nm}}(k)]^H = [\mathbf{y_{nm}}]^H\mathbf{V}(k)^H
    \left(\mathbf{V}(k) \mathbf{V}(k)^H + \frac{\sigma_n^2}{\sigma_s^2}\mathbf{I}\right)^{-1}
\label{eq:cnm=ynmV^-1}
\end{equation}
The regularization by the term $\frac{\sigma_n^2}{\sigma_s^2}\mathbf{I}$ helps stabilizing the matrix inversion. This solution, when substituted into Eq. (\ref{eq:anm_hat=Casmx}), yields the estimated Ambisonics coefficients, $\hat{a}_{nm}(k)$.
\color{black}
Note that the ASM filter is derived based on the assumption of $Q$ equal-energy, uncorrelated plane waves, similar to a diffuse-field model. Nevertheless, the authors of~\cite{BSM_journal_paper} showed that when the linear encoder is ideal, i.e. leading to zero encoding error, the encoding error remains zero for any sound field or source configuration.  This behavior is also relevant for the ASM solution, as it shares a formulation similar to ~\cite{BSM_journal_paper}. However, further study of this issue is beyond the scope of this paper and is proposed for future work. 
\color{black}

Eq. (\ref{eq:cnm=ynmV^-1}) provides a formulation of the ASM filter for each order and degree, $n,m$. Combining all filters into one large matrix will be useful for further development, and is given by:
\begin{equation}
\begin{split}
    [\mathbf{C}_\text{ASM}(k)]^H &= [\mathbf{c}_{00}(k),\dots,\mathbf{c}_{N_aN_a}(k)]^H \\
    &= \mathbf{Y}_{\Omega_Q}^H\mathbf{V}(k)^H
    \left(\mathbf{V}(k) \mathbf{V}(k)^H + \frac{\sigma_n^2}{\sigma_s^2}\mathbf{I}\right)^{-1}
    \label{eq:CASM=[c00,..,cNN]=YV^-1}
\end{split}
\end{equation}
Binaural reproduction using encoded Ambisonics via the ASM filter can be performed using (\ref{eq:p=hnmTanm}), replacing the Ambisonics vector $\mathbf{a_{nm}}(k)$ with its estimate from (\ref{eq:anm_hat=Casmx}):
\begin{equation}
    \hat{p}^{l,r}_{\text{ASM}}(k) = [\mathbf{h}_{\mathbf{nm}}^{l,r}(k)]^T \big[\tilde{\mathbf{C}}_\text{ASM}(k)\big]^H  \mathbf{x}(k)
    \label{eq:pASM=hnmCnmAMBx}
\end{equation}
where $\hat{p}^{l,r}_{\text{ASM}}(k)$ represents the binaural signal reproduced using the ASM filter, and $\tilde{\mathbf{C}}_\text{ASM}(k) = [\tilde{\mathbf{c}}_{00}(k), ...,\tilde{\mathbf{c}}_{N_aN_a}(k)]$ denotes the sign-inverted and index-rearranged ASM filter, with $\tilde{\mathbf{c}}_{nm}(k) = (-1)^m\mathbf{c}_{n,-m}(k)$.

\section{theoretical Limits of Performance}
\label{section:theoretical Limits of Performance}

In this section, we explore the limitations of the signal-independent method described in Sec. \ref{section: Ambisonics Encoding from Arbitrary Arrays}, aiming to understand the mathematical underpinnings that limit its performance.

\subsection{Limits on the Number of Ambionics Channels}

From the formulation of the error in (\ref{eq:Ambisonics analytical error}), several key factors affecting the accuracy of Ambisonics reconstruction through ASM can be identified. Neglecting noise, accurate reconstruction requires the following condition:
\begin{equation}
    \mathbf{V}(k)^H\mathbf{c_{nm}}(k) = \mathbf{y_{nm}}
    \label{eq:Vcnm=ynm}
\end{equation}
This indicates that accurate reconstruction depends on the accurate representation of $\mathbf{y_{nm}}$ using a linear combination of the columns of $\mathbf{V}(k)^H$, or in other words, on the projection of $\mathbf{y_{nm}}$ onto the null space of $\mathbf{V}(k)^H$ being zero. Given that the spherical harmonics vectors $\mathbf{y_{nm}}$ are generally orthogonal for all $0 \leq n$ and $-n \leq m \leq n$ over the direction space \cite{SH_Processing-book}, and that the steering matrix, with dimensions $M \times Q$, has a rank of at most $M$ because typically \mbox{$M\ll Q$}, the upper limit for the number of accurately encoded Ambisonics channels is:
\begin{equation}
    \# \text{AMB} \leq M
    \label{cond:AMB<M}
\end{equation}
where $\# \text{AMB}$ represents the number of encoded Ambisonics channels for each wave length $k$. The condition in (\ref{cond:AMB<M}) generalizes the condition in (\ref{cond:(Na+1)^2<M}), which specifically applies to uniformly distributed spherical arrays.

Note that, this condition only limits the number of encoded channels, leaving the specific channels that can be encoded undefined.

\subsection{Limit on the Accuracy of Ambisonics Channels}
\label{section:Limit on the Accuracy of Ambisonics Channels}

The reconstruction of Ambisonics channels is fundamentally tied to the characteristics of the steering matrix, which is dependent on the number and position of microphones, as well as the physical structure they are mounted on. To evaluate the accuracy of Ambisonics reconstruction, this work incorporates the error measurement technique proposed in \cite{ASM}. This method provides a systematic way to assess how effectively the encoded channels capture the desired spatial information.

The error metric used is given by:
\begin{equation}
    \begin{aligned}
        \xi_{\text{null}} = 10 \log_{10} \left( \frac{\left\lVert\mathbf{V_0}(k)\mathbf{y_{nm}}\right\rVert_2^2}{\left\lVert\mathbf{y_{nm}}\right\rVert_2^2} \right) \le \textnormal{TH},
    \end{aligned}
    \label{cond:yV0V0y<th}
\end{equation}
where $\mathbf{V_0}(k)$ is the null space of $\mathbf{V}(k)^H$, derived from the singular value decomposition (SVD) of $\mathbf{V}(k)^H$, specifically constructed from the eigenvectors associated with sufficiently small eigenvalues \cite{SVD}. Here, $\mathbf{V}(k)^H$ is a matrix of size $Q \times M$ (assuming $M < Q$), and the dimensions of $\mathbf{V_0}(k)$ are lower bounded by $Q \times (Q-M)$. Following the approach in \cite{ASM}, the threshold $\text{TH}$ is selected as $-10\ \text{dB}$, which ensures effective reconstruction within an acceptable error margin.

The factors influencing the condition in (\ref{cond:yV0V0y<th}) are primarily related to microphone placement and wavelength. Properly spaced microphones lead to increased spatial variability, which reduces dependencies between the microphone steering vectors. This minimizes the contributions of $\mathbf{V_0}(k)$ and improves reconstruction accuracy. Conversely, as frequency decreases and wavelength increases, spatial variability diminishes, increasing dependencies between the steering vectors. This expands $\mathbf{V_0}(k)$ and results in reduced encoding accuracy.

\subsection{Limits on the Magnitude of Ambisonics Encoding}
\label{section:Limits on the Magnitude of Ambisonics Encoding}

While the Ambisonics encoding error as defined in (\ref{eq:Ambisonics analytical error}) is a useful measure of performance, when this error is high, it may be useful to investigate the magnitude of the encoded Ambisonics signals, to understand whether these high error originate from magnitude differences. In particular, ASM filter that cannot accurately project the steering matrix to the desired spherical harmonics, as in (\ref{eq:Ambisonics analytical error}), tend to attenuate such projection to avoid high errors.

To quantify the attenuation of the ASM-encoded Ambisonics, the noise term in (\ref{eq:error_nm}) is neglected. Under this condition, it becomes evident that the ASM filters approximate the spherical harmonics by $\mathbf{V}(k)^H\mathbf{c}_{nm}(k)$. 

Consequently, the magnitude of the approximation is given by, $\big\Vert \mathbf{c}_{nm}^H(k) \mathbf{V}(k) \big\Vert_2^2$, while the ideal spherical harmonic magnitude is $\big\Vert \mathbf{y_{nm}}  \big\Vert_2^2$. 
Thus, the \color{black} Log Spectral Error (LSE) \color{black} can be formulated as:
\begin{equation}
\xi_{nm,\text{LSE}} = 10\text{log}_{10}\bigg(\frac{\big\Vert \mathbf{c}_{nm}^H(k) \mathbf{V}(k) \big\Vert_2^2}{\big\Vert \mathbf{y_{nm}}  \big\Vert_2^2} \bigg)
\label{cond:effective magnitude}
\end{equation}
with $\xi_{nm,\text{LSE}}[dB]$ ideally approaching $0\,$dB.

\color{black}

\section{Proposed Method for Array-Aware Encoding}
\label{section:Proposed Method for binaural reproduction}

This section presents a novel method for binaural reproduction for array-encoded Ambisonics, addressing the limitations of the encoding process. The proposed approach leverages array-specific preprocessing, incorporating a MagLS HRTF tailored to the array geometry, to enhance binaural reproduction quality.
Rewriting Eq. (\ref{eq:pASM=hnmCnmAMBx}) and omitting the dependence on $k$ for simplicity, we get
\begin{equation}
    \hat{p}_{\text{ASM}}^{l,r} = \big[\mathbf{h}^{l,r}_{\mathbf{nm}}\big]^{T} \big[\tilde{\mathbf{C}}_{\text{ASM}}\big]^H \mathbf{x}
    \label{eq:p_hat=hnmCASMx}
\end{equation}
where $\hat{p}_{\text{ASM}}^{l,r}$ denotes the reproduced binaural signal for the left or right ears.

The MSE between this estimated binaural signal and a reference binaural signal, $p^{l,r}$ can be written as:
\begin{equation}
    \begin{aligned}
        \scalebox{1.5}{\(\varepsilon\)}_{\text{Bin}} = E\left[\left\lVert \hat{p}_{\text{ASM}}^{l,r} - p^{l,r}\right\rVert_2 ^2\right]\bigg/E\left[\left\lVert p^{l,r} \right\rVert_2 ^2\right]
    \end{aligned}
    \label{eq:error_bin}
\end{equation}
Substituting the encoded Ambisonics signal from (\ref{eq:p_hat=hnmCASMx}) and the microphone signal from (\ref{eq:mic model}) into the error expression (\ref{eq:error_bin}), \color{black} neglecting the noise term \color{black}, and adopting the assumptions made in (\ref{eq:Ambisonics analytical error}), namely, that $\mathbf{s}$ exhibits a spatial correlation structure consistent with a diffuse sound field, the resulting error becomes:
\begin{equation}
        \scalebox{1.5}{\(\varepsilon\)}_{\text{Bin}} = \sigma_s^2\left\lVert [\mathbf{h}^{l,r}_{\mathbf{nm}}]^T\tilde{\mathbf{C}}_{\text{ASM}}^H\mathbf{V} - [\mathbf{h}^{l,r}]^T\right\rVert_2^2
    \label{eq:epsilon_Bin_with_R-ASM}
\end{equation}
By taking the absolute value of each term in (\ref{eq:epsilon_Bin_with_R-ASM}), the complex error expression is replaced with a magnitude-based error, yielding:
\begin{equation}
    \scalebox{1.5}{\(\varepsilon\)}_{\text{Bin}}^{\text{AA-MagLS}} = \sigma_s^2\left\lVert \left|[\mathbf{h}^{l,r}_{\mathbf{nm}}]^T\tilde{\mathbf{C}}_{\text{ASM}}^H\mathbf{V}\right| - \left|[\mathbf{h}^{l,r}]^T\right|\right\rVert_2^2
    \label{eq:epsilon_Bin_with_R-ASM^Mag}
\end{equation}
Similar to (\ref{eq:epsilon_MagLS}), this formulation represents the magnitude error in binaural reproduction using the Ambisonics format; however, (\ref{eq:epsilon_Bin_with_R-ASM^Mag}) employs encoded Ambisonics channels in place of the ideal Ambisonics channels.

This formulation provides a means of assessing the magnitude error in binaural reproduction via ASM at high frequencies. Moreover, it can serve as an array-aware objective function for optimizing the HRTF, with the goal of enhancing binaural reproduction quality.



This novel MagLS optimized HRTF is obtained by minimizing the following objective function:  
\begin{equation}
    \mathbf{h}^{l,r}_{\mathbf{nm}, \text{AA-MagLS}}(k) = \arg \min_{\mathbf{h}^{l,r}_{\mathbf{nm}}(k)} \varepsilon_{\text{Bin}}^{\text{AA-MagLS}}
    \label{eq:hnm_optimization}
\end{equation}
where $\mathbf{h}^{l,r}_{\mathbf{nm}, \text{AA-MagLS}}(k)$ denotes the proposed AA-MagLS HRTF.
Similar to (\ref{eq:argmin_epsilon_MagLS}), the use of the AA-MagLS HRTF is particularly relevant at high frequencies and is implemented by substituting it for the MagLS HRTF in (\ref{eq:hnm_weighted}).

Note that the optimization problem in (\ref{eq:hnm_optimization}) similar to (\ref{eq:argmin_epsilon_MagLS}) is non-convex, and standard iterative algorithms do not guarantee a globally optimal solution. Instead, they provide an approximate solution, which depends on the chosen initialization and optimization strategy. Techniques such as those described in \cite{book-Ambisonics} and \cite{Ambisonics_MagLS} can be employed to achieve practical solutions.

Also note that when ASM accurately reconstructs the Ambisonics channels, the MagLS HRTF and the proposed AA-MagLS HRTF become identical.
The ASM error in (\ref{eq:Ambisonics analytical error}), neglecting noise, equals zero when
\begin{equation}
    \mathbf{V}^H\mathbf{c}_{nm} = \mathbf{y}_{nm}
    \label{eq:Vcnm=ynm}
\end{equation}
$\forall 0 \le n \le N_a, -n \le m \le n$.
Thus, by considering all channels, (\ref{eq:Vcnm=ynm}) can be concatenated into:
\begin{equation}
    \mathbf{V}^H\mathbf{C}_{\text{ASM}} = \mathbf{Y}_{\Omega_Q}
    \label{eq:VCasm=Y}
\end{equation}
Now, by utilizing the complex conjugate property of spherical harmonics \cite{SH_Processing-book},
\begin{equation}
    Y_{nm}(\theta,\phi) = (-1)^m[Y_{n,-m}(\theta,\phi)]^* 
    \label{eq:Ynm=(-1)^mYn-m}
\end{equation}
and applying the Hermitian operation, (\ref{eq:Vcnm=ynm}) becomes:
\begin{equation}
    [\tilde{\mathbf{C}}_{\text{ASM}}]^H\mathbf{V} = \mathbf{Y}_{\Omega_Q}^T
    \label{eq:CV=Y}
\end{equation}
Finally, (\ref{eq:CV=Y}), which holds when the ASM reconstruction is perfect, can be substituted into the AA-MagLS HRTF objective function in (\ref{eq:epsilon_Bin_with_R-ASM^Mag}). In this case, the formulation becomes mathematically equivalent to the MagLS HRTF objective in (\ref{eq:epsilon_MagLS}).
This equality establishes AA-MagLS as a generalization of the MagLS HRTF for wearable arrays.

\color{black}
\section{AA-MagLS Minimization Algorithm}
The MagLS optimization employed in this work follows an iterative procedure consistent with the mathematical formulation in Eqs. (4.57)–(4.59) of \cite{book-Ambisonics}, adapted here to the proposed AA-MagLS HRTF framework. The optimization problem to be solved is defined in Eq. (\ref{eq:epsilon_Bin_with_R-ASM^Mag}). In the first step, a relaxation is introduced to transform this MagLS problem into an LS one as in Eq. (\ref{eq:epsilon_Bin_with_R-ASM}) by taking the magnitude and adding phase to $[\mathbf{h}^{l,r}]^T$ as follows
\begin{equation}
\begin{aligned}
    [\mathbf{h}^{l,r}_{\text{AA-MagLS}}(k_i)]^T = &  |\mathbf{h}^{l,r}(k_i)|^T \odot e^{j\phi(k_{i-1})}
    \label{eq:relaxedH}
\end{aligned}
\end{equation}
where $\mathbf{h}^{l,r}_{\text{AA-MagLS}}$ replaces $\textbf{h}^{l,r}$ in (\ref{eq:epsilon_Bin_with_R-ASM}), and \color{black} $\odot$ denotes the element-wise (Hadamard) product. Note that the iterations are applied across discrete wavenumbers (frequencies) in ascending order, and so it is assumed that the solutions at the low frequencies are computed using LS rather than MagLS which is indeed the case with HRTFs \cite{book-Ambisonics}. Now the solution of Eq. (\ref{eq:epsilon_Bin_with_R-ASM}) using Eq. (\ref{eq:relaxedH}) leads to 
\begin{equation}
\begin{aligned}
    [\mathbf{h}^{l,r}_{nm, \text{AA-MagLS}}(k_i)]^T = & [\mathbf{h}^{l,r}_{\text{AA-MagLS}}(k_i)]^T 
    \left[ \tilde{\mathbf{C}}_{\text{ASM}}^H(k_i) \mathbf{V}(k_i) \right]^{\dagger}
    \label{eq:relaxesMagLS}
\end{aligned}
\end{equation}
where $[\cdot]^{\dagger}$ represents the pseudo-inverse operator. Note that here, unlike the standard MagLS approach, the steering matrix and the ASM filter are directly incorporated into the solution.

Finally, the phase vector $\phi(k_{i-1})$ is propagated from the outcome of the previous iteration's:
\begin{equation}
    \phi(k_{i-1}) = \angle \left( [\mathbf{h}^{l,r}_{nm, \text{AA-MagLS}}(k_{i-1})]^T \tilde{\mathbf{C}}_{\text{ASM}}^H(k_{i-1}) \mathbf{V}(k_{i-1}) \right)
    \label{eq:phaseMagLS}
\end{equation}
where $\angle(\cdot)$ denotes the element-wise phase operator. Note that the weighting process described in Eq. (\ref{eq:hnm_weighted}) is also applied here, i.e. taking a weighted average of the AA-MagLS solution from (\ref{eq:relaxesMagLS}) and the LS solution, using the frequency-dependent weighting factor $\alpha(k_i)$ defined in (\ref{eq:alpha}).

\color{black}
\section{Summary of the Proposed AA-MagLS Method vs. MagLS for HRTF Encoding}
The proposed method extends conventional HRTF preprocessing by introducing the AA-MagLS formulation for binaural reproduction. This approach incorporates array-specific knowledge into the error minimization process, allowing for better tailoring to imperfect or non-ideal Ambisonics encoding. The key elements of the AA-MagLS approach compared to the MagLS approach are outlined in  Table \ref{fig:table1}. These include aspects about the computation and use of the two methods. As can be seen, the AA-MagLS approach requires more information and may therefore be less standard but this may be balanced by improved performance for tailored arrays.

\begin{table}[h]
\centering
\normalsize
\caption{Comparison of MagLS and AA-MagLS for HRTF Encoding}
\renewcommand{\arraystretch}{1.2}
\begin{tabular}{@{}p{4cm}ll@{}}
\toprule
\text{Aspect} & \text{MagLS} & \text{AA-MagLS} \\ \midrule
1) Ambisonics encoder used in the computation& No & Yes \\[2.5ex]
 
2) Array steering function used in the computation & No & Yes \\[2.5ex]

3) Formulation for combining Ambisonics and HRTF & Standard & Standard \\[2.5ex]

4) Tailoring to imperfect Ambisonics encoding & No & Yes \\[2.5ex]

5) Use of off-the-shelf encoded HRTF & Yes & No \\
\bottomrule
\end{tabular}
\label{fig:table1}
\end{table}

\section{Simulation Study: Ambisonics Encoding Limitations}
\label{section:Simulation Study: Ambisonics Encoding Limitations}
This section provides a simulation-based evaluation of Ambisonics encoding using an array designed to resemble a wearable glasses microphone array. This evaluation assesses the array's limitations for Ambisonics encoding, based on the methodology outlined in Sec.\ref{section:Limit on the Accuracy of Ambisonics Channels}.

\subsection{Setup}
\label{section:Array Setup}
\begin{figure}[H]
    \centering
    \begin{subfigure}[b]{\columnwidth}
        \centering
        \includegraphics[width=\linewidth]{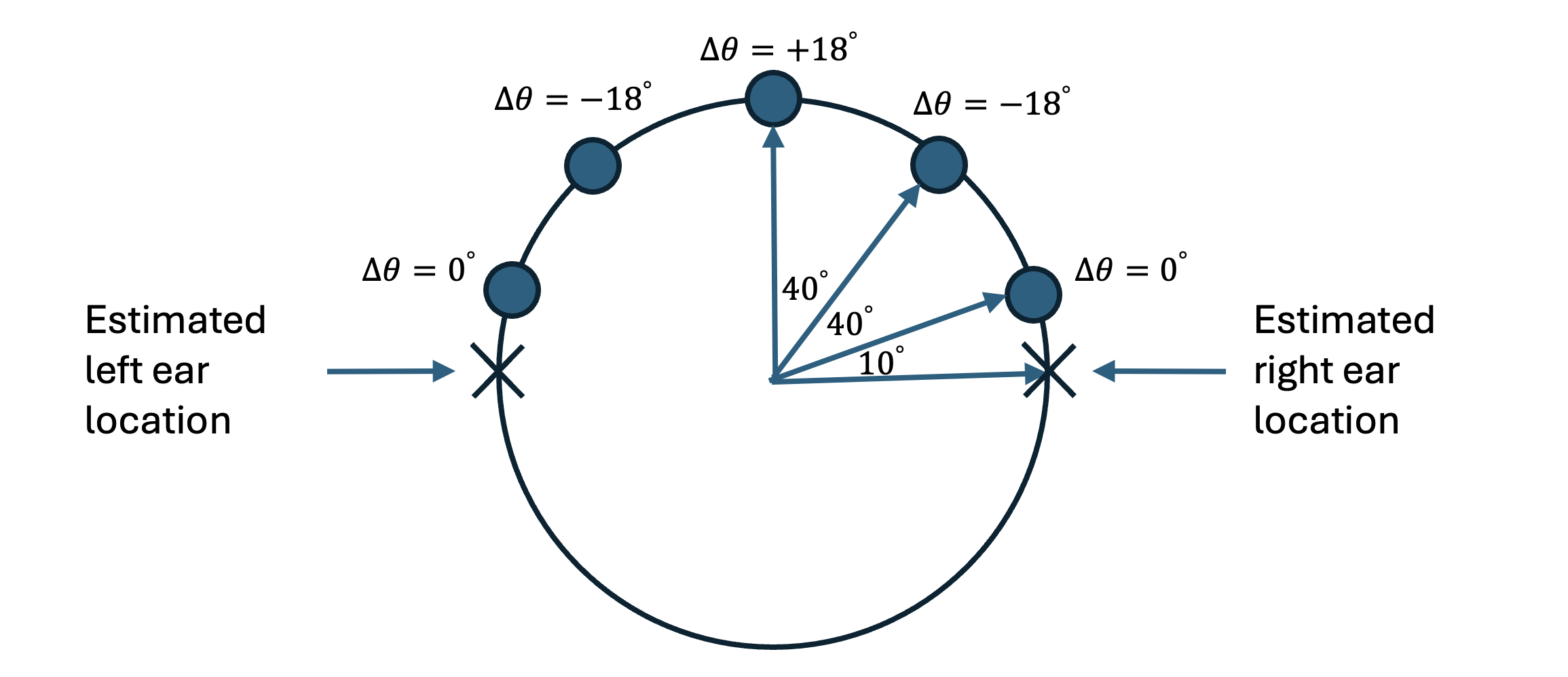}
        \caption{2D view}
        \label{fig:array_plot_2D}
    \end{subfigure}
    
    \vspace{1em} 

    \begin{subfigure}[b]{\columnwidth}
        \centering
        \includegraphics[width=0.7\linewidth]{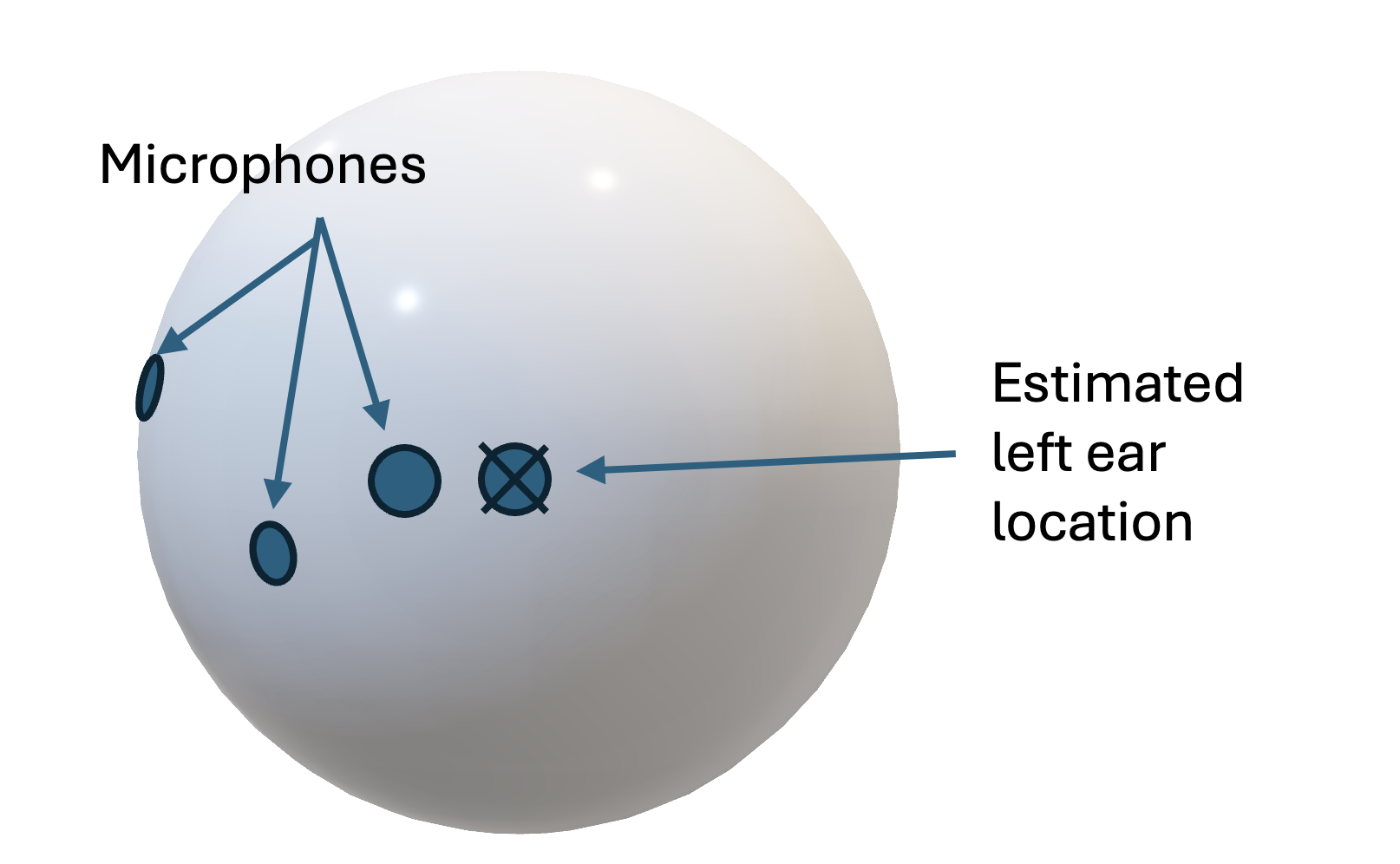}
        \caption{3D view}
        \label{fig:array_plot_3D}
    \end{subfigure}
    
    \caption{\fontsize{9}{11}\selectfont Microphone positions on a rigid sphere with locations $(\theta, \phi)$: \( \left\{ (90^{\circ}, -80^{\circ}), (72^{\circ}, -40^{\circ}), (108^{\circ}, 0^{\circ}), (72^{\circ}, 40^{\circ}), (90^{\circ}, 80^{\circ}) \right\} \). The estimated ear locations correspond to \( (90^{\circ}, \pm 90^{\circ}) \).}
    \label{fig:array_plots}
\end{figure}

The microphone array consists of \( M = 5 \) microphones arranged along a semi-circle with a radius of \( 0.1 \) m, mounted on a rigid sphere. The location of each microphone is defined in spherical coordinates \((\theta, \phi)\), where \(\theta\) represents the elevation angle and \(\phi\) represents the azimuth angle. The microphone positions are given by:  
\[
\left\{ (90^{\circ}, -80^{\circ}), (72^{\circ}, -40^{\circ}), (108^{\circ}, 0^{\circ}), (72^{\circ}, 40^{\circ}), (90^{\circ}, 80^{\circ}) \right\}
\]
as illustrated in Fig. \ref{fig:array_plots}.  
This configuration is designed to resemble a wearable glasses microphone array, providing a compact arrangement. The microphones are primarily positioned along the horizontal plane but also incorporates slight variations in elevation, with angles deviating by \(\pm 18^{\circ}\). This design choice allows the array to capture elevation-related acoustic cues, enhancing spatial encoding beyond purely horizontal localization.  

\subsection{Ambisonics Encoding Performance}
\label{section:Ambisonics Encoding}
To study the Ambisonics encoding accuracy with the simulated array, we employ the metric defined in (\ref{cond:yV0V0y<th}), which quantifies the projection of the SH basis functions onto the null space of the steering matrix. Fig.~\ref{fig:Ambisonics cond n=0,1,2} presents the resulting metric \( \scalebox{1.5}{\(\xi\)}_{\text{null}} \) (in dB), evaluated for different SH orders.
\begin{figure}
    \centering
    \includegraphics[width=0.45\textwidth]{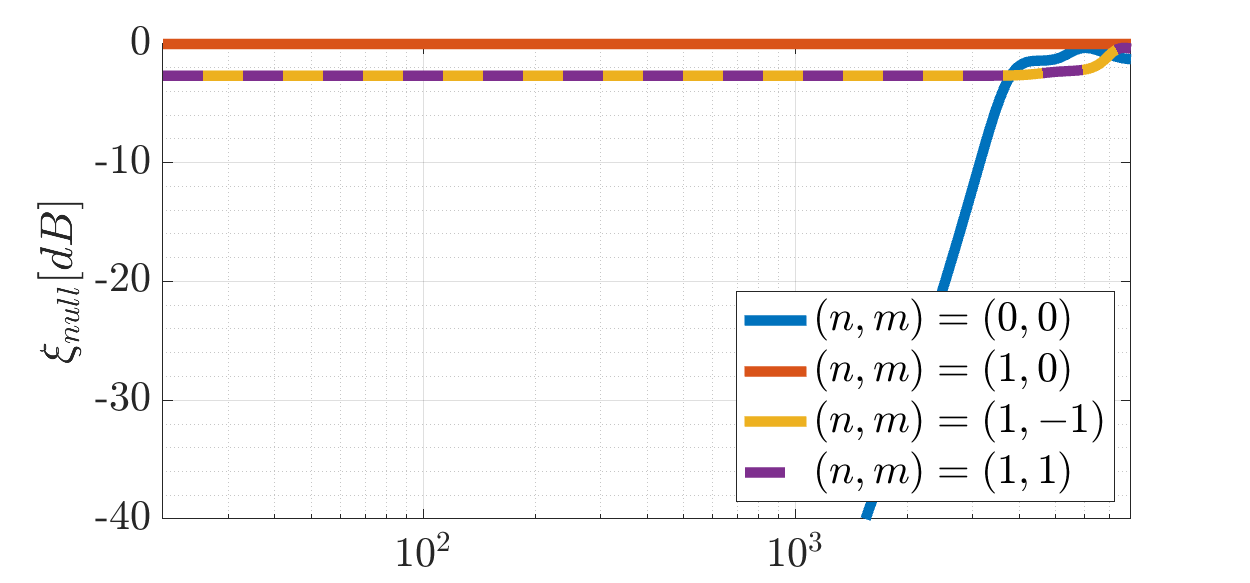}
    \includegraphics[width=0.45\textwidth]{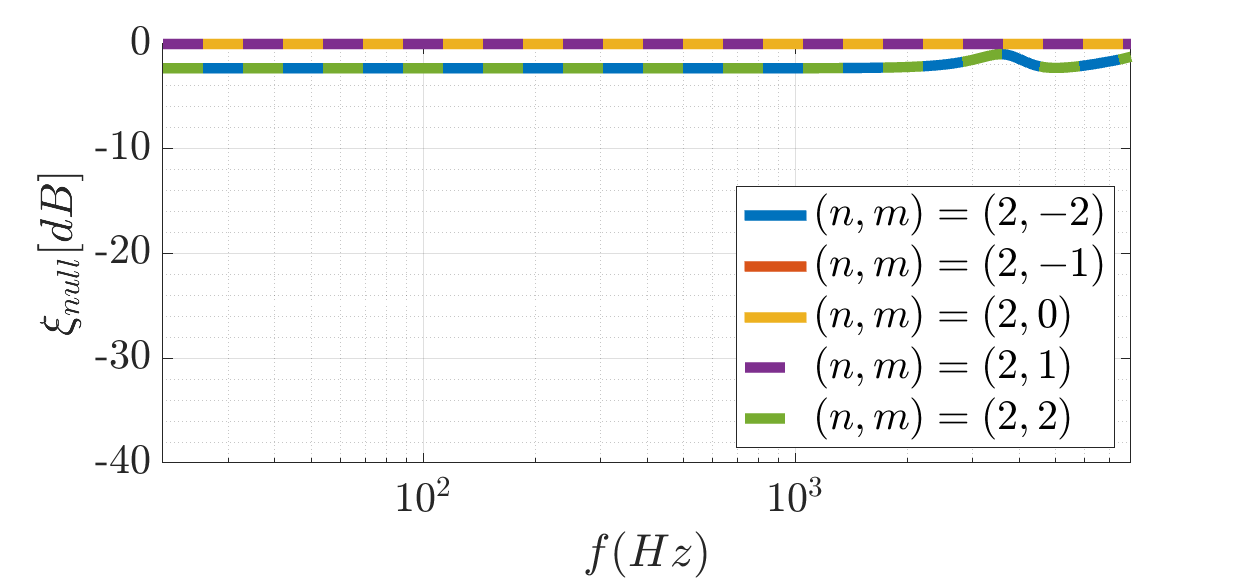}
    \caption{\fontsize{9}{11}\selectfont 
    The error \( \scalebox{1.5}{\(\xi\)}_{\text{null}} \) in dB, as defined in (\ref{cond:yV0V0y<th}), for the steering matrix of the array described in Sec.~\ref{section:Array Setup}. The evaluation is performed for two sets of SH orders. The top plot corresponds to SH orders \( (n,m) = (0,0), (1,-1), (1,0), (1,1) \), while the bottom plot represents \( (n,m) = (2,-2), (2,-1), (2,0), (2,1), (2,2) \).}
    \label{fig:Ambisonics cond n=0,1,2}
\end{figure}
The results presented in Fig.~\ref{fig:Ambisonics cond n=0,1,2} demonstrate that the proposed array is capable of encoding four Ambisonics channels corresponding to SH orders \( n=0 \) and \( n=1 \) with good accuracy at the low frequencies. Despite the array being primarily oriented in the horizontal plane, its slight variations in the vertical plane enable it to encode channels associated with altitude changes, such as \( (n,m) = (1,\pm1) \).
Furthermore, the plot reveals that four channels satisfy the condition defined in (\ref{cond:yV0V0y<th}), maintaining error levels below -10 dB across the low frequencies, particularly below $1\,$kHz. 
Fig.\ref{fig:Ambisonics cond n=0,1,2} also shows that SH of the second orders cannot be encoded accurately at all, emphasizing the limitation on the number of encoded SH as in (\ref{cond:yV0V0y<th}) for this case.

\subsection{Magnitude of Ambisonics Encoding}
As shown in the previous subsection, the ASM in the given example encodes Ambisonics channels with reasonable accuracy up to order $N_a=1$.
While the condition in (\ref{cond:yV0V0y<th}) as illustrated in Fig. \ref{fig:Ambisonics cond n=0,1,2} suggests successful reconstruction for frequencies below $1\,$kHz, it is evident that this measure increases at higher frequencies, indicating substantial reconstruction error.
To further characterize this error, we examine the LSE of the ASM using (\ref{cond:effective magnitude}) as detailed in Section~\ref{section:Limits on the Magnitude of Ambisonics Encoding}.

As illustrated in Fig. \ref{fig:ASM magnitude n=0,1}, the magnitude of the encoded Ambisonics channels is severely attenuated in the frequency range above $1\,$kHz precisely where the error becomes pronounced. This strong attenuation, reflecting the inability of the Encoding filters to project the steering matrix to the spherical harmonics functions, leading to diminishing magnitude of the filters, directly contributes to the substantial errors observed in the reconstructed Ambisonics channels.

\begin{figure}[H]
    \centering
    \includegraphics[width=0.45\textwidth]{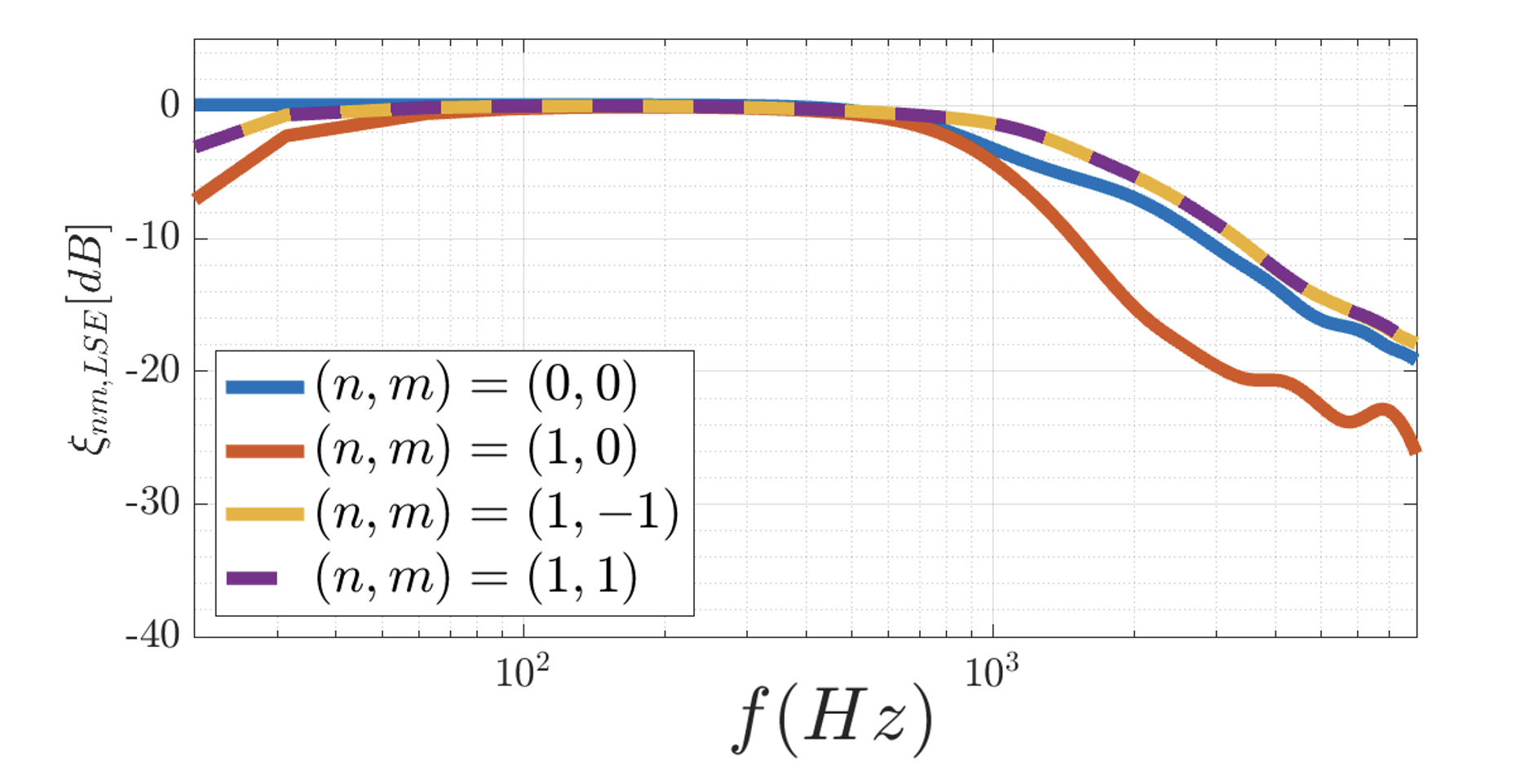}
    \caption{\fontsize{9}{11}\selectfont 
    The LSE of the ASM filter in dB, $\xi_{nm,\text{LSE}}$ for SH orders \( (n,m) = (0,0), (1,-1), (1,0), (1,1) \), as defined in (\ref{cond:effective magnitude}).The magnitude is evaluated for the array configuration described in Sec.~\ref{section:Array Setup}.}
    \label{fig:ASM magnitude n=0,1}
\end{figure}

\section{Simulation Study: Binaural Reproduction}
This section presents simulation-based experiments to evaluate the proposed AA-MagLS HRTF method for computing HRTF coefficients. Specifically, it examines the effect of using the AA-MagLS HRTF in conjunction with first-order Ambisonics encoded via ASM for binaural reproduction performance. The results are compared against the conventional MagLS HRTF approach.

\subsection{Experimental Setup}
\label{section:Experimental Setup}
The array used here is the same as the array presented in
Sec.~\ref{section:Array Setup}. The HRTFs used in this study are based on the Neumann $KU100$ manikin, taken from \cite{HRTF_data_set}, sampled using the Lebedev scheme \cite{LEBEDEV197610} with 2702 points, and transformed into the SH domain with a maximum order of \( N_h = 30 \). The
estimated ear locations are determined based on the HRTF
dataset, with the left and right ears positioned symmetrically
around the median plane as in Fig.\ref{fig:array_plots}.

\subsection{Methodology}
\label{section:Methodology}
The performance of the proposed AA-MagLS HRTF method for binaural reproduction is evaluated and compared to several other methods, as detailed in this section.

Each method used in this section consists of given Ambisonics encoding, followed by a convolution with an HRTF for binaural reproduction. Three approaches to compute the coefficients of the HRTF in the SH domain are applied depending on the method. These variants include HRTF MagLS \cite{HRTF_MagLS} by minimizing Eq. (\ref{eq:epsilon_MagLS}), denoted as \emph{MagLS HRTF}, the proposed array-dependent MagLS, minimizing Eq. (\ref{eq:epsilon_Bin_with_R-ASM^Mag}), denoted \emph{AA-MagLS HRTF}, and direct linear encoding, as in (\ref{eq:p=hnmTanm}), denoted \emph{regular HRTF}. For both \emph{MagLS HRTF} and \emph{proposed MagLS HRTF}, the MagLS formulations are applied only at high frequencies, as specified in (\ref{eq:alpha}). Here is an outline of the methods used in this section:
\begin{itemize}

    \item \textbf{HOA:}  
    High-Order Ambisonics (HOA) encoding up to order \( N=30 \), followed by binaural reproduction using a \emph{regular HRTF}.
    
    \item \textbf{ASM + MagLS HRTF:}  
    First-order ASM encoding is applied, followed by binaural reproduction using the \emph{MagLS HRTF}. 

    \item \textbf{ASM + AA-MagLS HRTF:}
    Similar to the previous method but utilizes the proposed \emph{AA-MagLS HRTF}.

\end{itemize}

\subsection{Binaural Reproduction Error}
\label{section: Binaural Reproduction Error}
To assess the performance of each binaural reproduction method, the binaural NMSE presented in (\ref{eq:epsilon_Bin_with_R-ASM}) is re-written here, neglecting the noise term:
\begin{equation}
        \scalebox{1.5}{\(\varepsilon\)}_{\text{Bin}} = \sigma_s^2\left\lVert [\mathbf{h}^{l,r}_{\mathbf{nm}}]^T\tilde{\mathbf{C}}_{\text{ASM}}^H\mathbf{V} - [\mathbf{h}^{l,r}]^T\right\rVert_2^2 \bigg/  \left\lVert [\mathbf{h}^{l,r}]^T \right\rVert_2^2 \label{eq:e=|hnmCASMV-h|^2/|h|^2}
\end{equation}

Now, to assess the magnitude error of the models, the following absolute value terms are incorporated into (\ref{eq:e=|hnmCASMV-h|^2/|h|^2}):
\begin{equation}
    \scalebox{1.5}{\(\varepsilon\)}_{\text{Bin}}^{\text{Mag}} = \left\lVert [|\mathbf{h}^{l,r}_{\mathbf{nm}}]^T\tilde{\mathbf{C}}_{\text{ASM}}^H\mathbf{V}| - |[\mathbf{h}^{l,r}]^T|\right\rVert_2^2 \bigg/ \left\lVert  [\mathbf{h}^{l,r}]^T \right\rVert_2^2
    \label{eq:eMAG=|hnmCASMV-h|^2/|h|^2}
\end{equation}
Since the models aim to minimize both the MSE and magnitude MSE across different frequencies, the combined error is introduced as a function of frequency $f$ in Hz:
\begin{equation}
    \scalebox{1.5}{\(\varepsilon\)}_{\text{Bin}}^{\text{comb}}(f) = (1 - \alpha(f)) \scalebox{1.5}{\(\varepsilon\)}_{\text{Bin}}(f) + \alpha(f) \scalebox{1.5}{\(\varepsilon\)}_{\text{Bin}}^{\text{Mag}}(f)
    \label{eq:e=(1-a)e+ae^Mag}
\end{equation}
where $\scalebox{1.5}{\(\varepsilon\)}_{\text{Bin}}$ and $\scalebox{1.5}{\(\varepsilon\)}_{\text{Bin}}^{\text{Mag}}$ correspond to the left and right ear errors of (\ref{eq:e=|hnmCASMV-h|^2/|h|^2}) and (\ref{eq:eMAG=|hnmCASMV-h|^2/|h|^2}), and $\alpha(f)$ is defined as in (\ref{eq:alpha}).
To assess performance with head rotation compensation, the MSE in (\ref{eq:e=|hnmCASMV-h|^2/|h|^2}) is modified for the rotated versions of ASM using the Wigner-D matrix \cite{Wigner-D}. 

\begin{figure}
    \centering
    \includegraphics[width=0.45\textwidth]{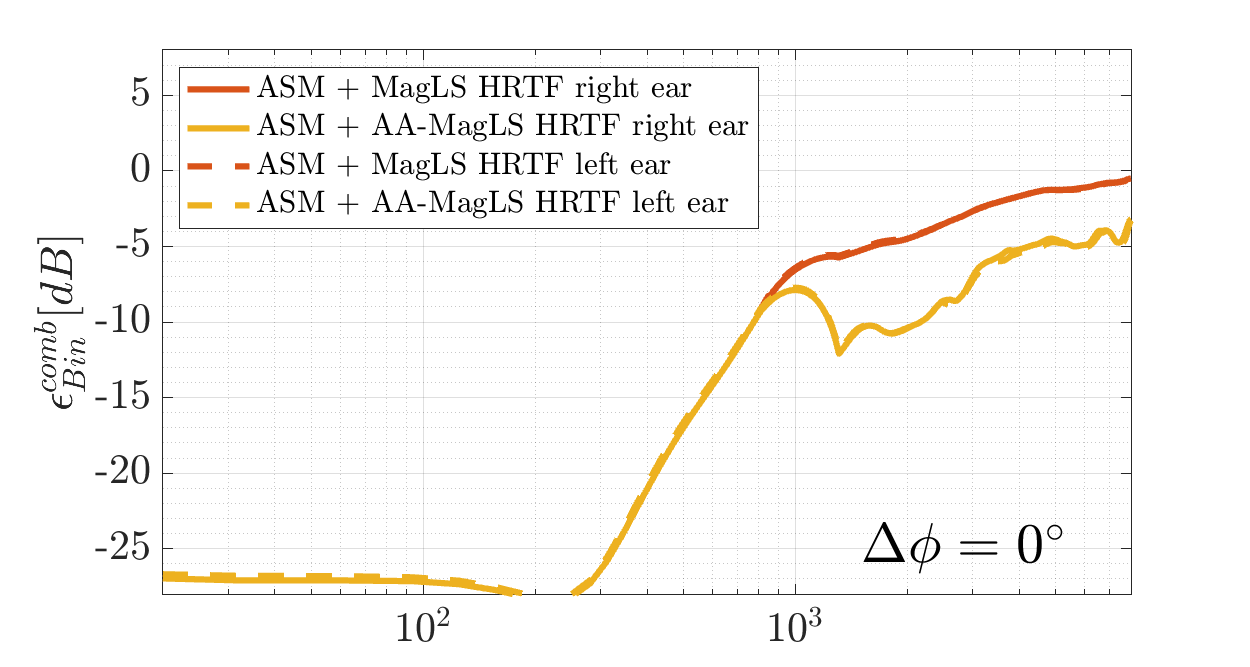}
    \includegraphics[width=0.45\textwidth]{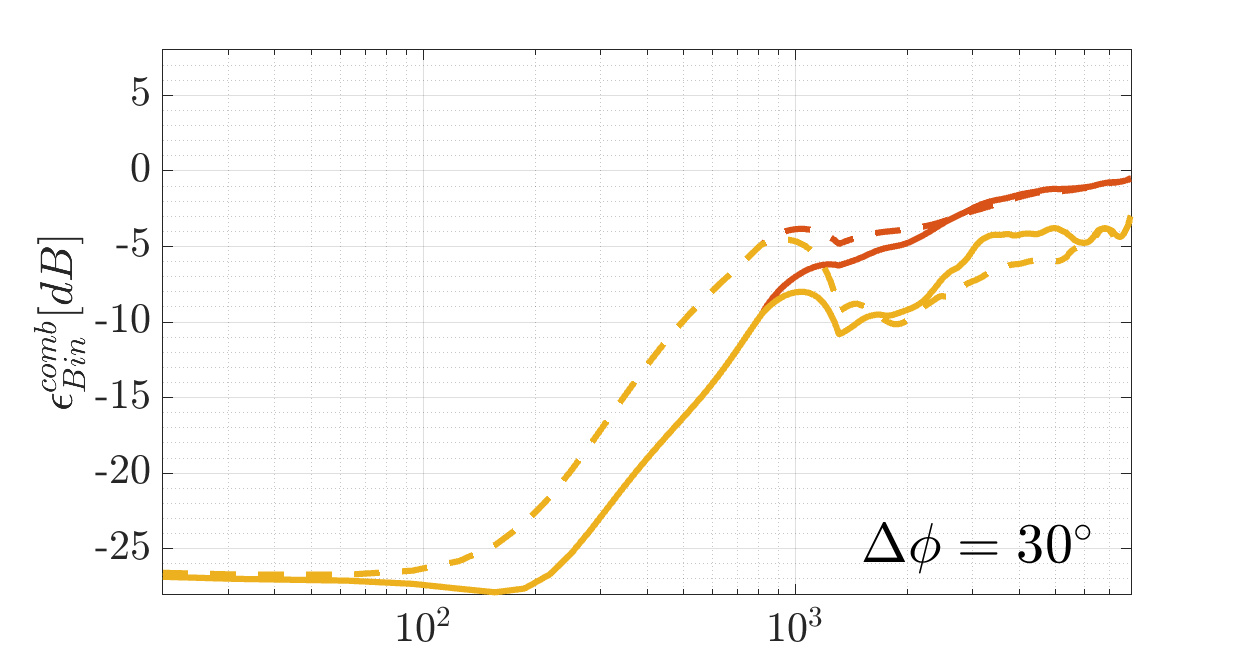}
    \includegraphics[width=0.45\textwidth]{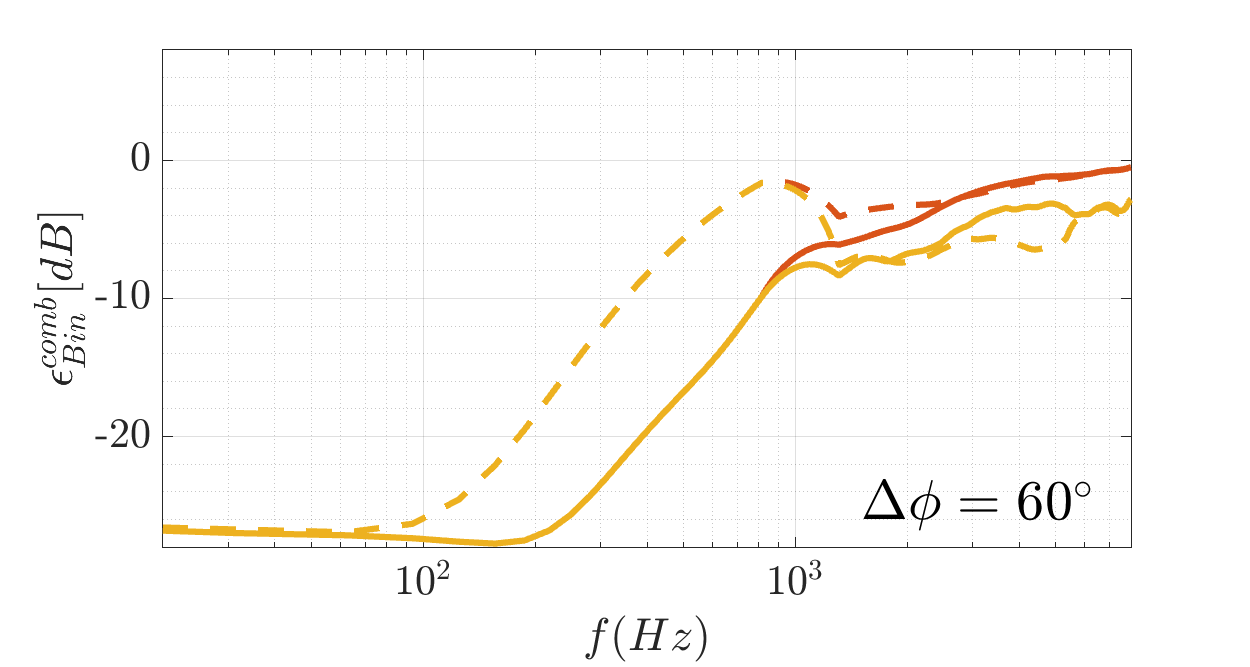}
    \caption{Measure of the error in (\ref{eq:e=(1-a)e+ae^Mag}) for \emph{ASM + MagLS HRTF}, \emph{ASM + proposed MagLS HRTF}, for both ears. The results are shown with azimuthal head rotations of \(0^\circ\), \(30^\circ\), and \(60^\circ\) from top to bottom, where the rotation is applied using Wigner-D functions.}
    \label{fig:HRTF Comparison with rotation}
\end{figure}

As can be seen in Fig.~\ref{fig:HRTF Comparison with rotation}, ASM + AA-MagLS HRTF outperforms ASM + MagLS HRTF in the high frequency regions where Magnitude-least square is applied. This trend remains consistent across the tested head orientations.
However, for larger head rotation angle, the overall error grows for both methods.

\subsection{ITD and ILD-Based Evaluation}
\label{section:ITD and ILD-Based Perceptual Evaluation Setup}

To complement the MSE-based evaluation presented in previous subsection, which may not fully capture perceptual aspects of binaural reproduction, we present a perceptually-motivated analysis based on interaural time difference (ITD) and interaural level difference (ILD).

\vspace{5mm}
\subsubsection{Setup}
The setup follows the methodology described in Sec.\ref{section:Methodology}, and is aligned with established perceptual metrics for evaluating binaural localization.

The binaural signals are generated in response to a sound field consisting of a single plane wave with a DOA defined by elevation \( \theta = 90^{\circ} \) and azimuth \( \phi \in [0^{\circ}, 359^{\circ}] \), sampled at \( 1^{\circ} \) resolution. The simulation parameters are consistent with those used in Sec. \ref{section:Simulation Study: Ambisonics Encoding Limitations}.

\subsubsection{Methodology for ITD and ILD analysis}
For ITD estimation, we employ the cross-correlation method proposed in~\cite{ITD-Andreopoulou2017}, which has been shown to reflect ITD cues. Prior to cross-correlation, the binaural signals are low-pass filtered with a cutoff frequency of $3\,$kHz, as recommended in ~\cite{ITD-Andreopoulou2017}. The interaural cross-correlation (IACC) is computed as:
\begin{equation}
    \text{IACC}_p(\tau) = \sum_{t=0}^{T - \tau - 1} p^l(t + \tau) \, p^r(t),
\end{equation}
where \( p^l(t) \) and \( p^r(t) \) are the left and right ear Head Related Impulse Response (HRIR) based signals and \( T \) is the total number of samples. The ITD is then estimated as:
\begin{equation}
    \text{ITD}(\theta, \phi) = \arg\max_{\tau} \{ \text{IACC}_p(\tau) \},
\label{eq:ITD}
\end{equation}
and compared to the reference ITD obtained using the original HRIRs, yielding the ITD error:
\begin{equation}
    \varepsilon_{\text{ITD}}(\theta, \phi) = \left| \text{ITD}(\theta, \phi) - \text{ITD}_{\text{ref}}(\theta, \phi) \right|.
    \label{eq:ITD error}
\end{equation}

ILD values are evaluated for each azimuth angle \( \phi \) using an energetic approach based on auditory filterbanks according to~\cite{ILD-Xie2013}. Specifically, we employ a bank of equivalent rectangular bandwidth (ERB) filters spanning the frequency range \([20, 8000]\)~Hz. Each left and right ear impulse response, denoted \( x_l(t, \phi) \) and \( x_r(t, \phi) \), is filtered through the ERB filterbank to model frequency-dependent loudness perception. For each filter with center frequency \( f_c(i) \), the filtered power spectra are computed, and the energy in each band is obtained as:
\begin{equation}
E^{l,r}(f_c(i), \theta, \phi) = \sum_{f = f_{\text{low}}}^{f_{\text{high}}} |H_i(f)| \cdot |X^{l,r}(f, \theta, \phi)|^2,
\end{equation}
where \( H_i(f) \) denotes the magnitude response of the \(i\)-th ERB filter, and \( |X^{l,r}(f, \phi)|^2 \) represent the power spectra of the filtered left and right signals at azimuth \( \phi \).

The ILD at each center frequency and azimuth is then computed as:
\begin{equation}
\text{ILD}(f_c(i), \theta, \phi) = 10 \cdot \log_{10} \left( \frac{E^l(f_c(i), \theta,\phi)}{E^r(f_c(i), \theta, \phi)} \right).
\end{equation}
This process produces ILD values across azimuth and frequency, reflecting perceptually meaningful spatial differences in levels between the two ears, thereby enabling a comprehensive evaluation of binaural reproduction fidelity. These values are then averaged over frequency, resulting in:
\begin{equation}
\text{ILD}(\theta, \phi) = \frac{1}{I}\sum_{i=1}^{I} \text{ILD}(f_c(i), \theta, \phi),
\label{eq:ILD}
\end{equation}
where $I$ is the number of frequency bands.
Finally, similar to (\ref{eq:ITD error}), the ILD error is computed as:
\begin{equation}
\varepsilon_{\text{ILD}}(\theta, \phi) = \left| \text{ILD}(\theta, \phi) - \text{ILD}_{\text{ref}}(\theta, \phi) \right|.
\label{eq:ILD_error}
\end{equation}

\subsubsection{Results}
\begin{figure*}[p]
\centering

\newcommand{\ColumnHeightTwo}{20cm} 

\begin{minipage}[t][\ColumnHeightTwo][t]{0.47\textwidth}
    \centering
    \includegraphics[width=\textwidth]{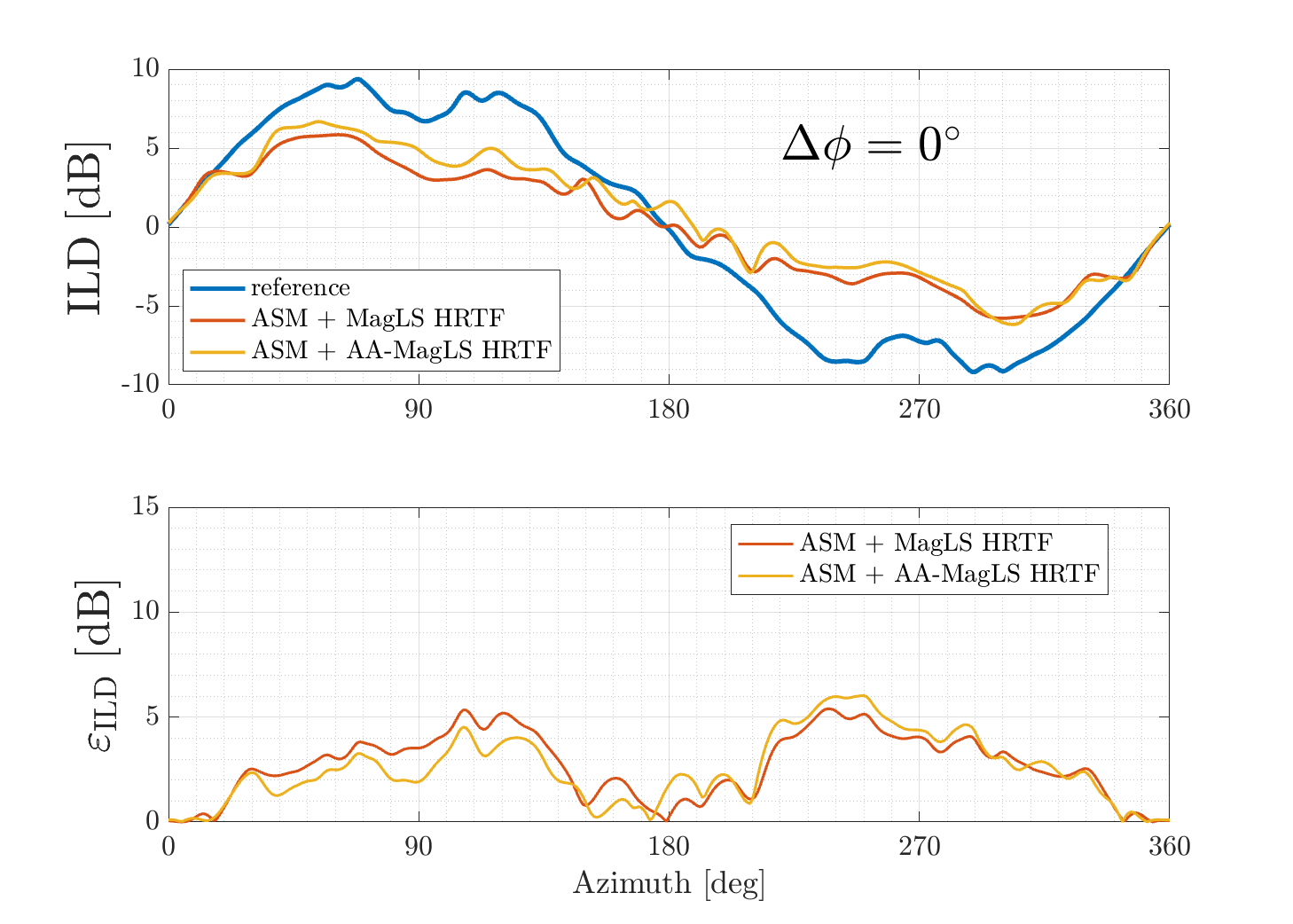}\\[4mm]
    \includegraphics[width=\textwidth]{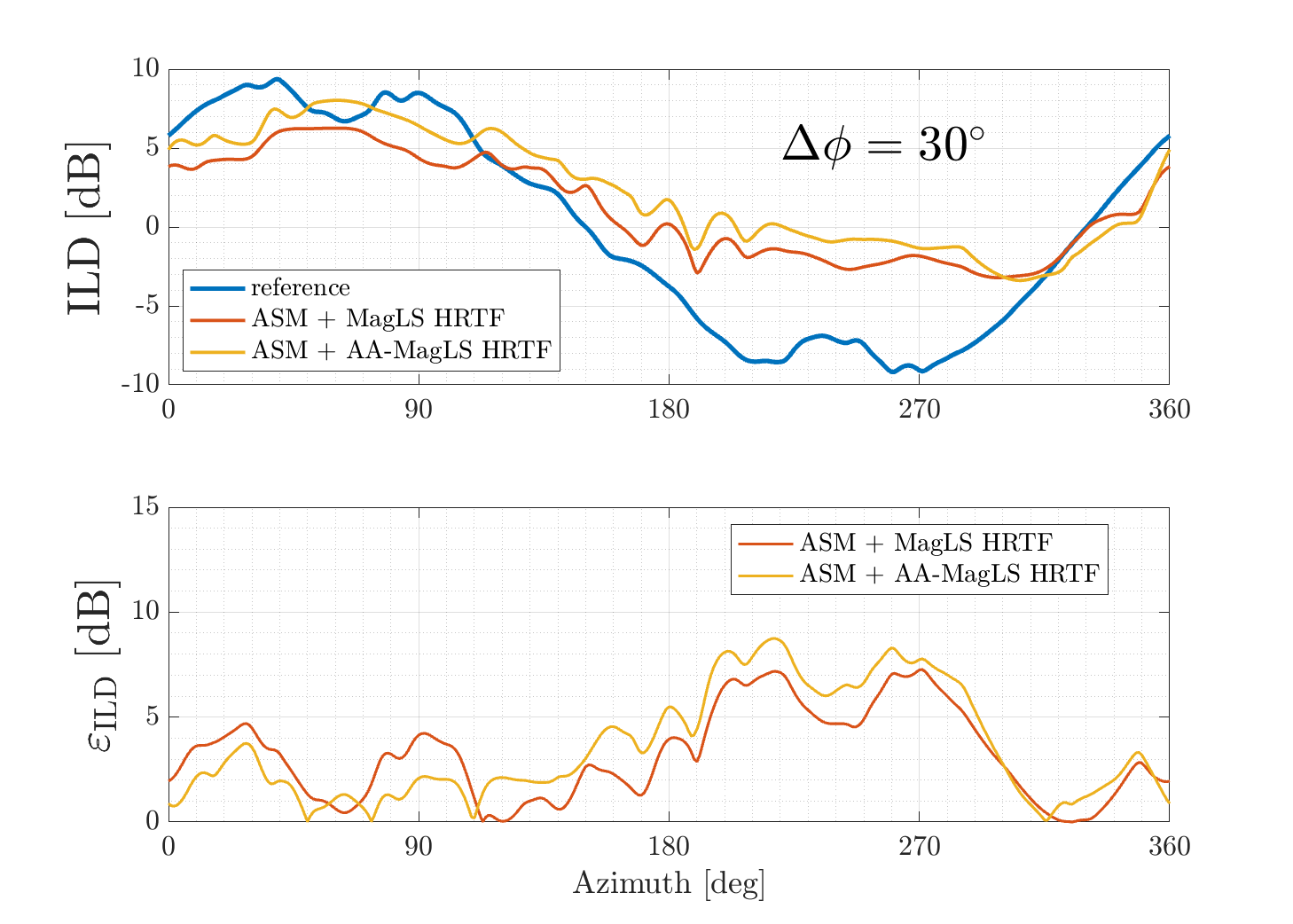}\\[4mm]
    \includegraphics[width=\textwidth]{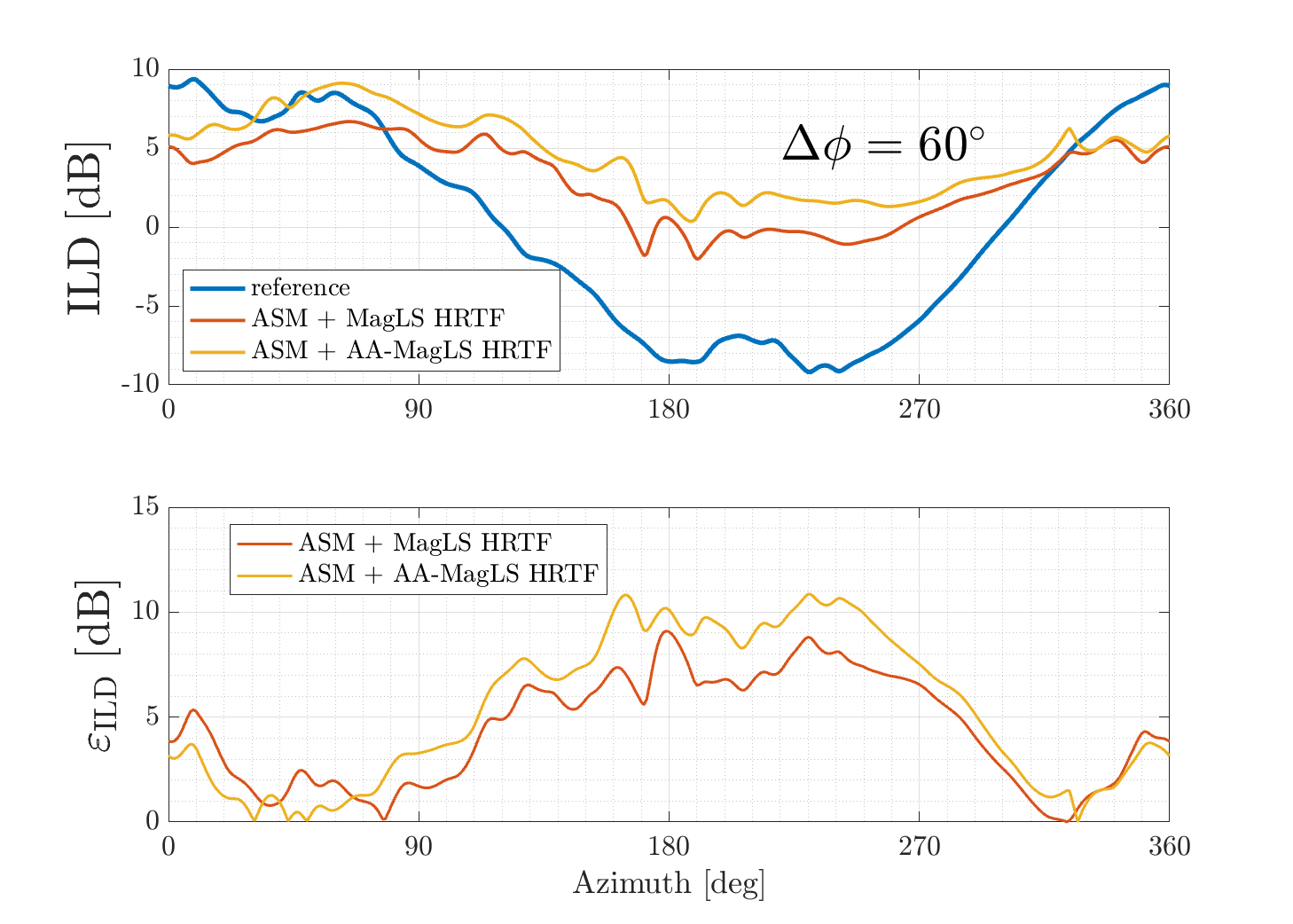}

    \caption{ILD as in (\ref{eq:ILD}) and ILD error as in (\ref{eq:ILD_error}) 
    for \emph{ASM + MagLS HRTF} and \emph{ASM + proposed MagLS HRTF}. 
    Results shown for head rotations of \(0^\circ\), \(30^\circ\), and \(60^\circ\).}
    \label{fig:ILD}
\end{minipage}%
\hfill
\begin{minipage}[t][\ColumnHeightTwo][t]{0.47\textwidth}
    \centering
    \includegraphics[width=\textwidth]{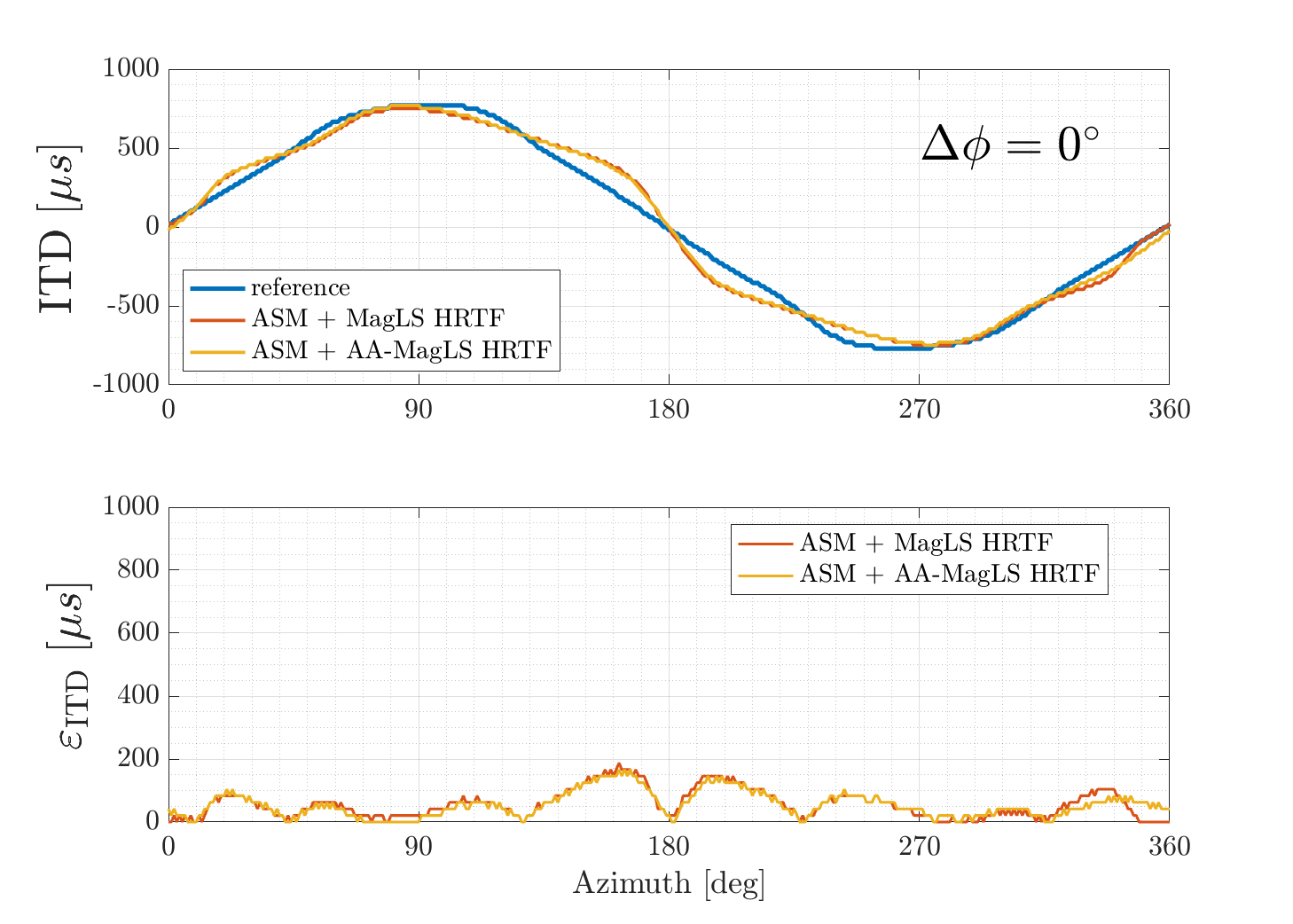}\\[4mm]
    \includegraphics[width=\textwidth]{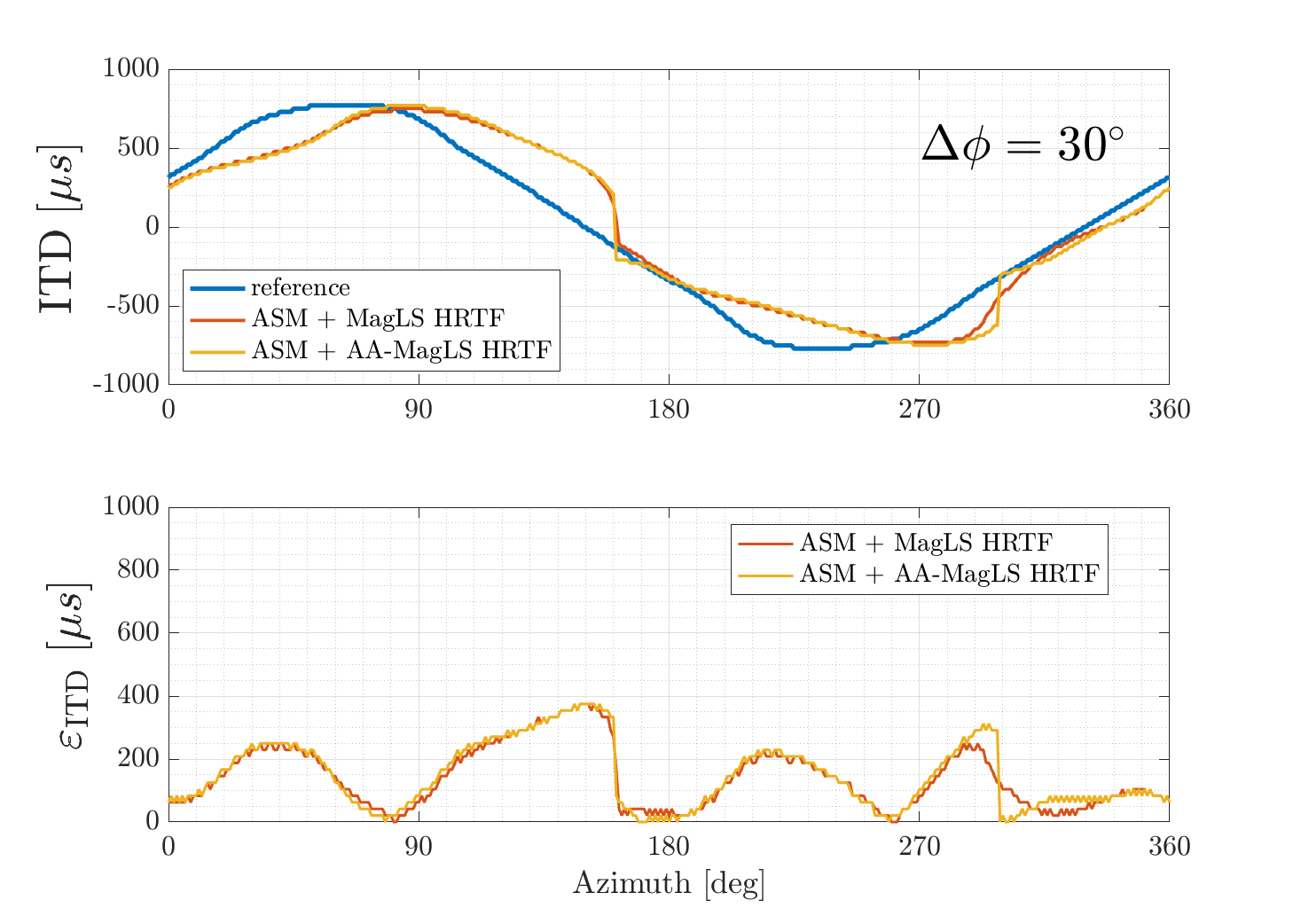}\\[4mm]
    \includegraphics[width=\textwidth]{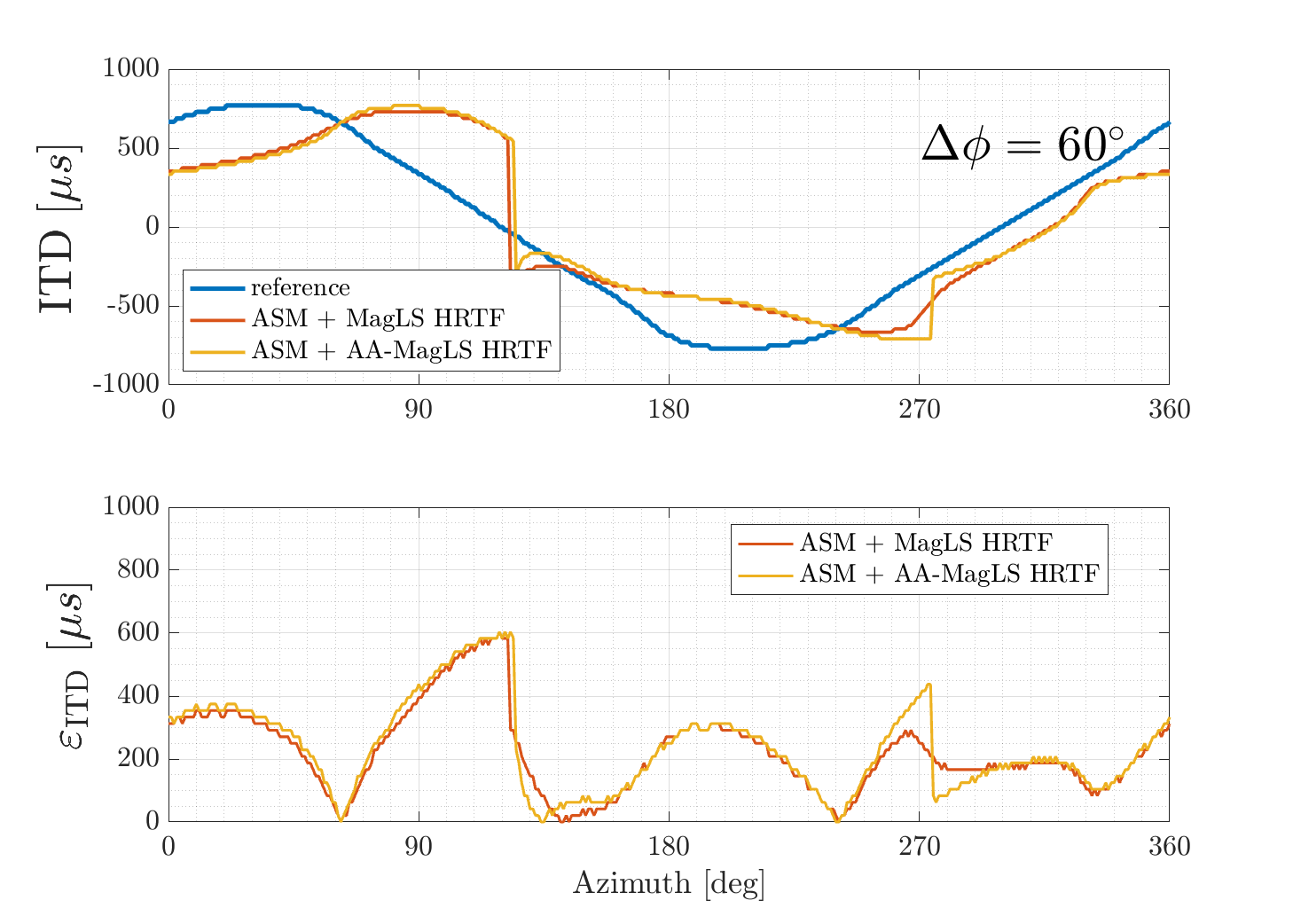}

    \caption{ITD as in (\ref{eq:ITD}) and ITD error as in (\ref{eq:ILD_error}) 
    for \emph{ASM + MagLS HRTF} and \emph{ASM + proposed MagLS HRTF}. 
    Results shown for head rotations of \(0^\circ\), \(30^\circ\), and \(60^\circ\).}
    \label{fig:ITD}
\end{minipage}

\end{figure*}

Fig.~\ref{fig:ILD} and Fig.~\ref{fig:ITD} present the ILD and ITD error analyses, respectively.  
Both evaluations were conducted under three head orientations: \(0^\circ\), \(30^\circ\), and \(60^\circ\).  
ASM + AA-MagLS and conventional ASM + MagLS exhibit similar ITD values error across all angles, which is expected as the ITD is dominant at low frequencies, where both approaches are identical.
In the ILD results, both methods show consistent error trends, with varying peaks and dips across azimuth angle, but no clear advantage for either approach.  
Overall, the ILD and ITD evaluations indicate that both methods perform similarly.

\color{black}
\section{Simulation Study: Effect of Array Configuration on the Benefit of AA-MagLS HRTF}
\label{section:Effect of array configuration}
The previous sections demonstrated that the AA-MagLS HRTF approach provided improved binaural reproduction compared with the more standard MagLS HRTF formulation for the array shown in Fig.~\ref{fig:array_plot_2D}. However, the degree of improvement may depend on the configuration of the microphone array used to encode the Ambisonics signals. Since AA-MagLS explicitly incorporates the array steering function into the HRTF optimization, its benefit over MagLS HRTF may depend on the steering function and therefore on array configuration.
To investigate this dependency, we compare two complementary metrics: the \emph{LSE} of the encoded Ambisonics filters, defined in (\ref{cond:effective magnitude}), and the resulting \emph{binaural error}, defined in (\ref{eq:e=(1-a)e+ae^Mag}).
The LSE characterizes how the array geometry shapes each encoded channel over frequency, whereas the binaural error reflects the perceptual consequence of these encoding distortions after employing AA-MagLS.
Evaluating both metrics together provides a direct link between an array’s encoding limitations and the achievable improvement obtained by the proposed array-aware HRTF processing.
\subsection{Setup}
To explore how array configuration influences AA-MagLS performance, four microphone arrays were selected. They differ in radius, number of microphones, and spatial distribution, thereby covering a broad spectrum of encoding conditions. The four microphone arrays, two of which are spherical, one linear and one circular, are detailed in Table \ref{table:arrays}.

\begin{table*}[t]
\centering
\color{black}
\begin{tabular}{|l|p{11cm}|}
\hline
\textbf{Array Name} & \textbf{Description} \\
\hline
\textbf{(a) Small Spherical Array} & Twenty microphones are uniformly distributed across the surface of a rigid sphere with a $0.03$ m radius. \\
\hline
\textbf{(b) Large Spherical Array} & Four microphones are uniformly distributed across the surface of a rigid sphere with a $0.043$ m radius.  \\
\hline
\textbf{(c) Linear Array} & Four microphones are placed along a linear array at the following positions along the x-axis: $-0.075$m, $-0.025$m, $0.025$m, and $0.075$m. \\
\hline
\textbf{(d) Circular Array} & Ten microphones are uniformly distributed along a circular array on the surface of a rigid sphere with a $0.15$ m radius. \\
\hline
\end{tabular}
\caption{\textcolor{black}{Details of the microphone arrays in the simulation study of Sec. \ref{section:Effect of array configuration}. The table describes arrays geometry and microphones numbers and positions.}}
\label{table:arrays}
\end{table*}

\subsection{Methodology}
The LSE (\ref{cond:effective magnitude}) and binaural error (\ref{eq:e=(1-a)e+ae^Mag}) were computed for all four arrays, using HRTF from (\ref{section:Experimental Setup}) and simulated steering matrices.

\subsection{Results}
\begin{figure*}[p]
\centering

\newcommand{\ColumnHeight}{22cm}

\begin{minipage}[t][\ColumnHeight][t]{0.47\textwidth}
    \centering
    \includegraphics[width=\textwidth]{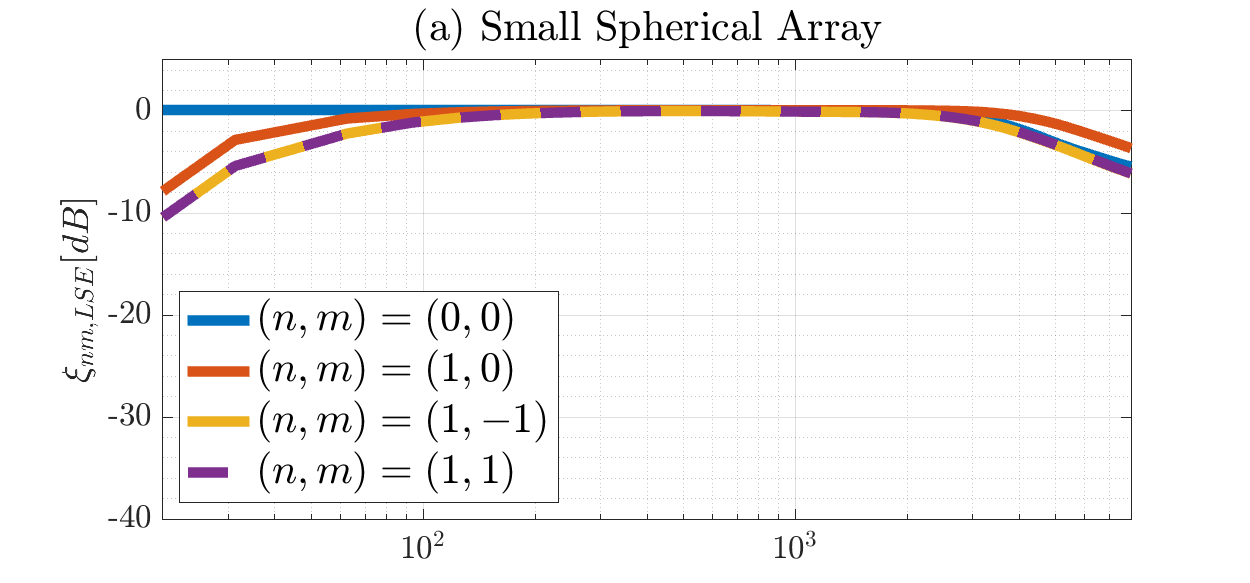}\\[6.2mm]
    \includegraphics[width=\textwidth]{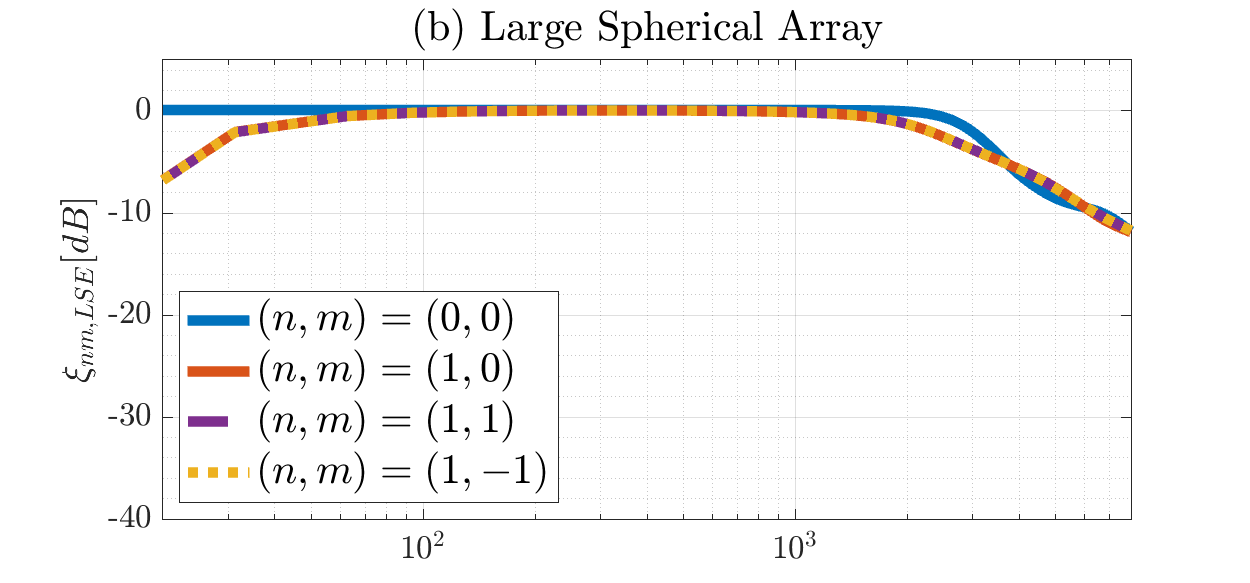}\\[7.6mm]
    \includegraphics[width=\textwidth]{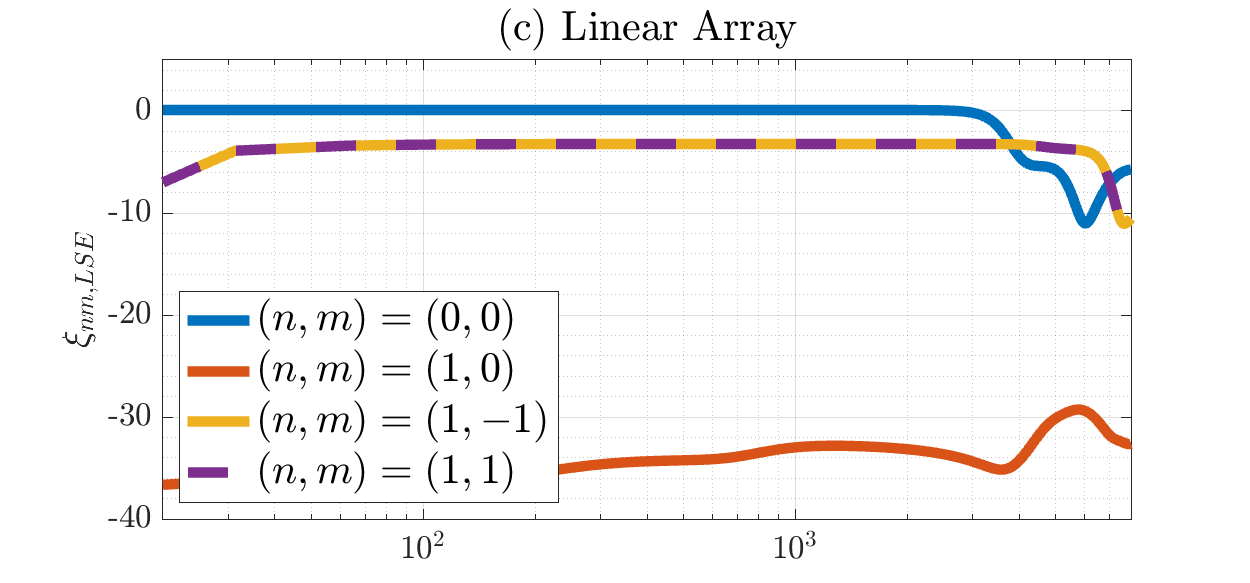}\\[8.2mm]
    \includegraphics[width=\textwidth]{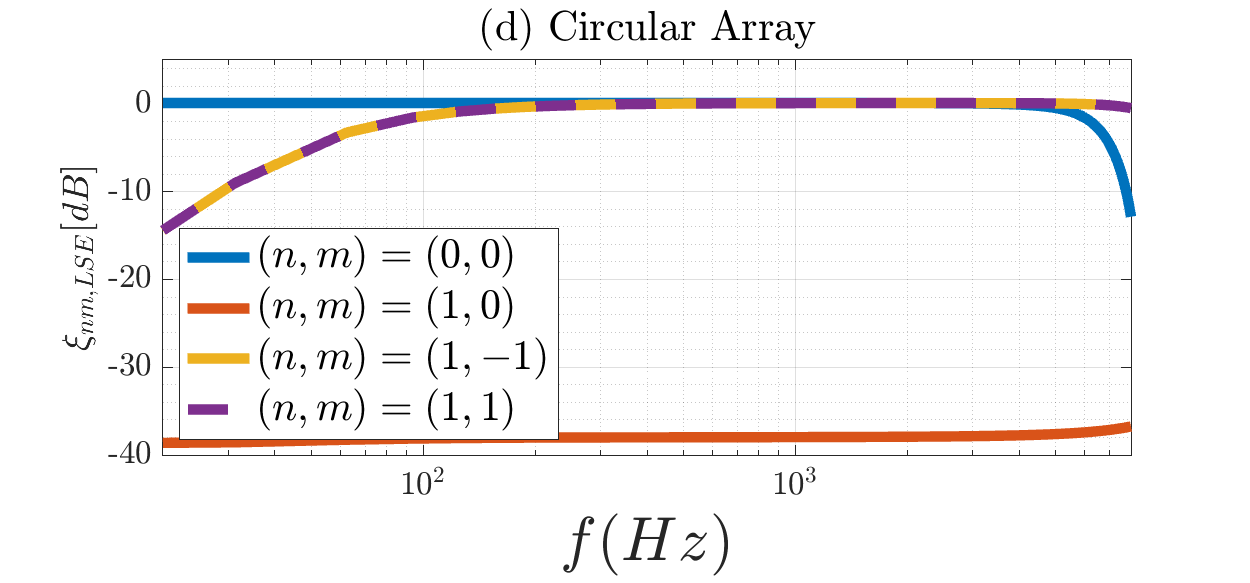}\\[2mm]
    \caption{\fontsize{9}{11}\selectfont \textcolor{black}{
    The LSE of the ASM filter in dB, $\xi_{nm,\text{LSE}}$ for SH orders \( (n,m) = (0,0), (1,-1), (1,0), (1,1) \), as defined in (\ref{cond:effective magnitude}). The magnitude is evaluated for arrays (a)-(d) as described in Sec.~\ref{section:Effect of array configuration} and Table \ref{table:arrays} from top to bottom.}}
    \label{fig:ASM_Mag_over_manyarrays}
\end{minipage}
\hfill
\begin{minipage}[t][\ColumnHeight][t]{0.47\textwidth}
    \centering
    \includegraphics[width=\textwidth]{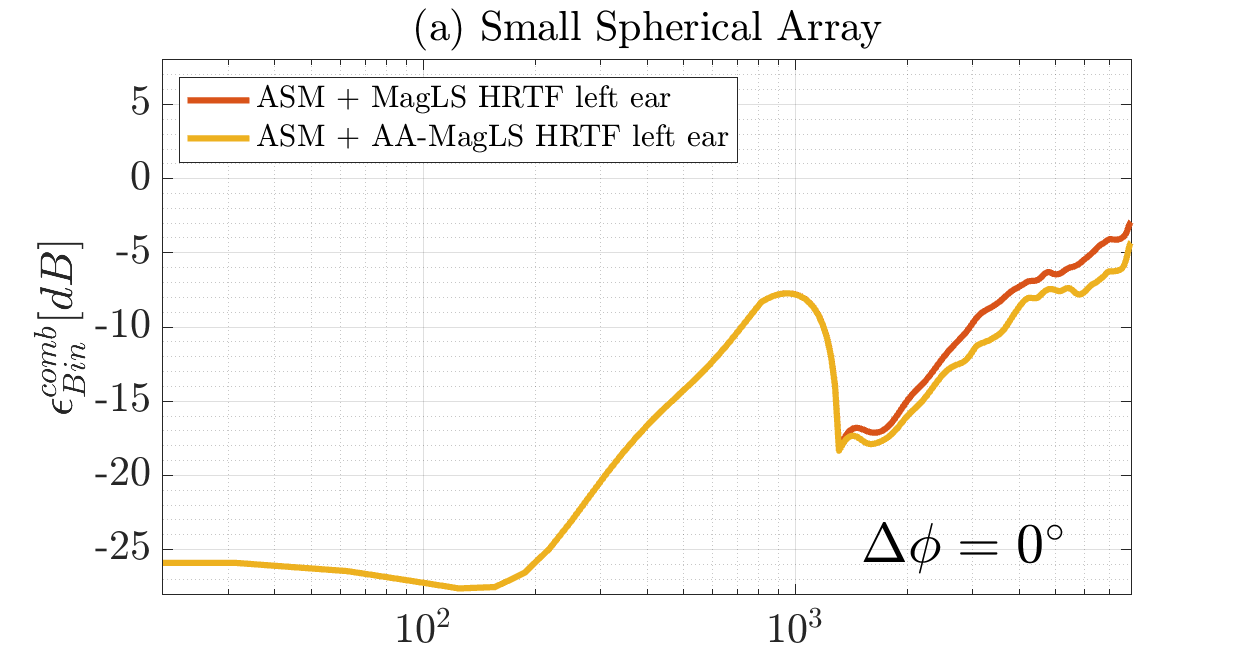}\\[2mm]
    \includegraphics[width=\textwidth]{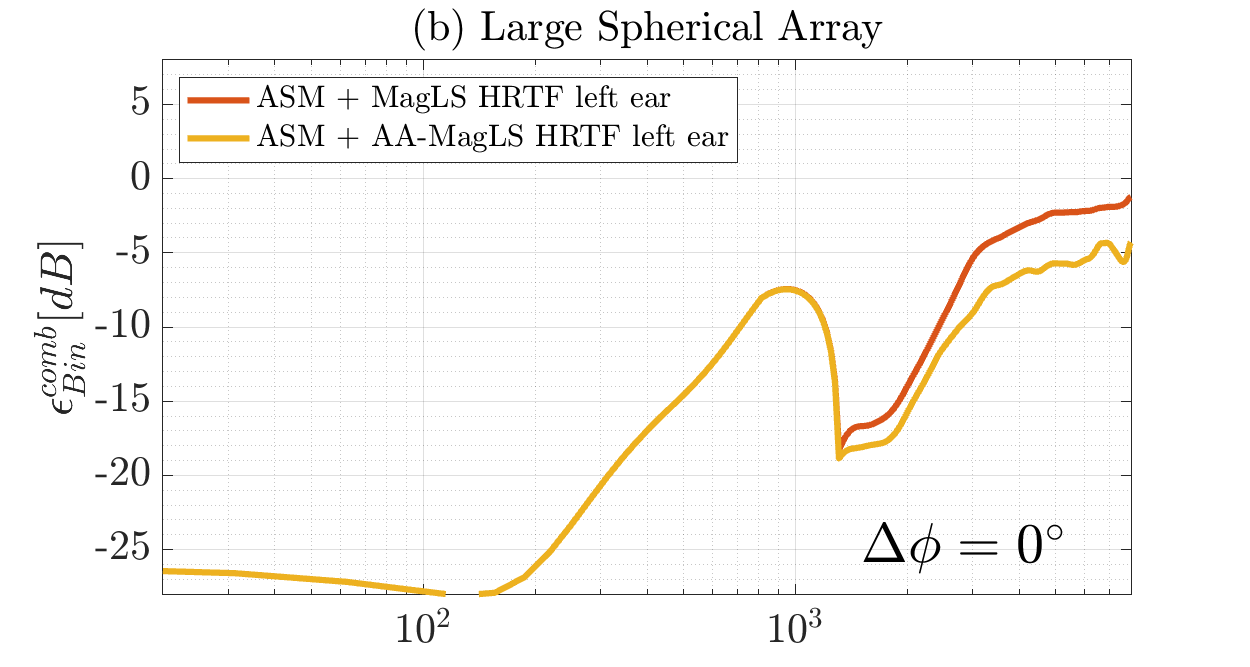}\\[2mm]
    \includegraphics[width=\textwidth]{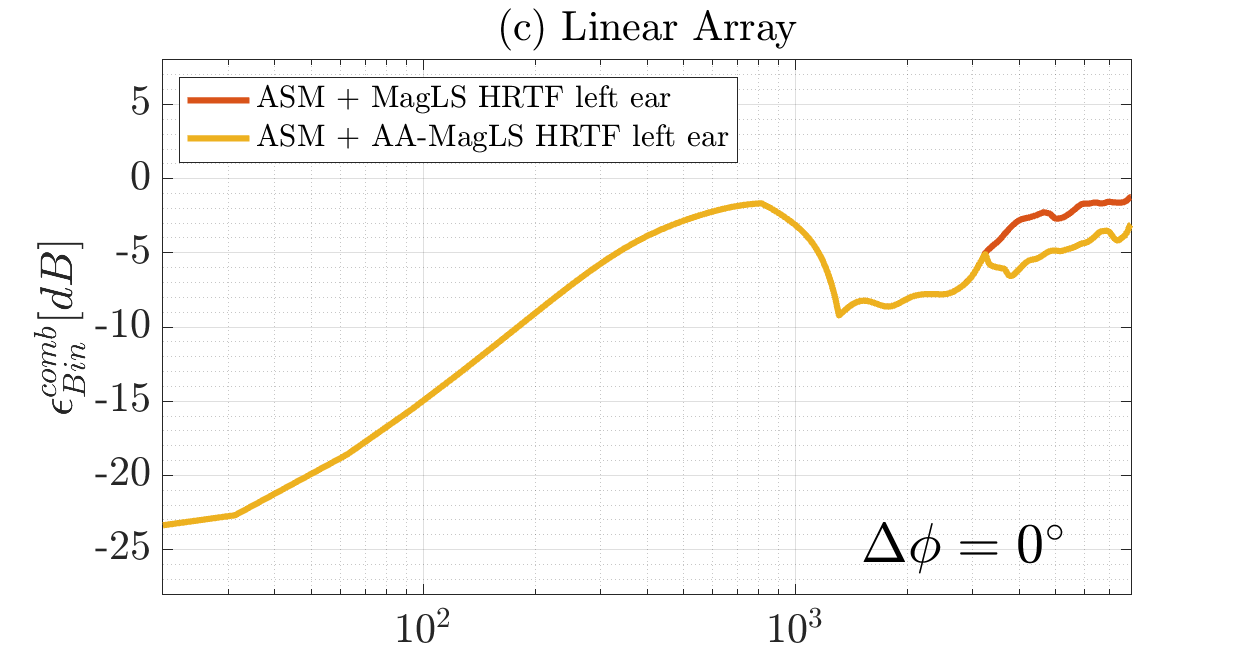}\\[2mm]
    \includegraphics[width=\textwidth]{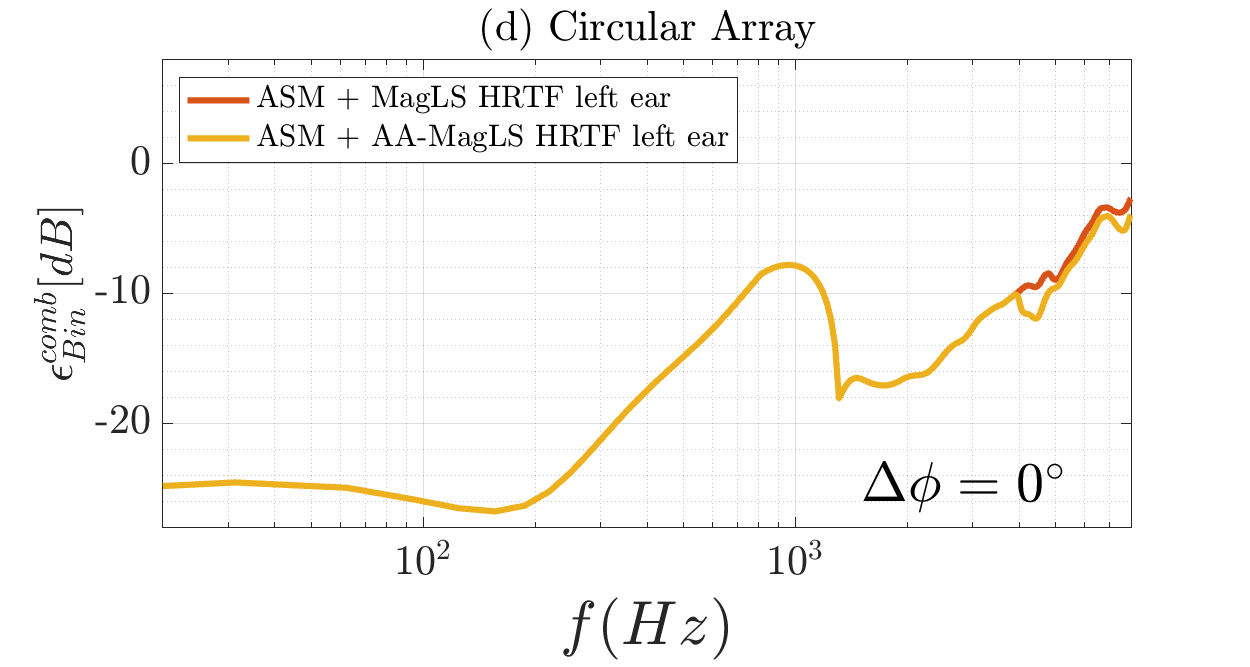}
    \caption{\textcolor{black}{Measure of the error in (\ref{eq:e=(1-a)e+ae^Mag}) for \emph{ASM + MagLS HRTF}, \emph{ASM + AA-MagLS HRTF}, shown for the left ear for convenience, results for the right ear are similar. The results are shown without azimuthal head rotation. The error is evaluated for the arrays (a)-(d) configurations as described in Sec.~\ref{section:Effect of array configuration} and Table \ref{table:arrays}  from top to bottom.}}
    \label{fig:bin_error_over_manyarrays}
\end{minipage}

\end{figure*}

The LSE of the encoded FOA channels for Arrays (a)--(d) are shown in Fig.~\ref{fig:ASM_Mag_over_manyarrays}, and the corresponding binaural errors obtained with MagLS and AA-MagLS HRTFs are presented in Fig.~\ref{fig:bin_error_over_manyarrays}. Taken together, these results reveal how the array geometry governs the potential improvement achievable by the proposed AA-MagLS HRTF formulation.

For Arrays (a) and (b), the LSEs of the FOA channels remain close to the ideal value across most of the frequency range, indicating accurate Ambisonics encoding with minimal array-induced attenuation. Under these conditions, the AA-MagLS HRTF provides little or no improvement over MagLS HRTF, as reflected in their nearly overlapping binaural-error curves for Array (a) and for Array (b) up to approximately 3\,kHz. Above this frequency, Array (b) exhibits increasing attenuation in all FOA channels, and AA-MagLS becomes beneficial, yielding a noticeable reduction in binaural error relative to MagLS.

Arrays (c) and (d), both restricted to horizontal microphone placement, show severe attenuation in the \((n,m)=(1,0)\) component due to the absence of vertical spatial aperture. For Array (c), additional attenuation appears in the remaining FOA channels above approximately 3.2\,kHz, and AA-MagLS improves the binaural error only in this frequency region. In contrast, Array (d) preserves the horizontal FOA channels with high accuracy and exhibits negligible differences between MagLS and AA-MagLS across the entire band, aside from the fundamentally unrecoverable \((n,m)=(1,0)\) mode.

Overall, the results demonstrate that the benefit of AA-MagLS depends on the nature of the encoding error introduced by the array geometry. When channels are accurately encoded, AA-MagLS offers no improvement. When channels are completely unobservable, as in the \((n,m)=(1,0)\) component for horizontal arrays, the method cannot recover the missing information. The approach is most advantageous in the intermediate case where channels exhibit moderate encoding error: sufficient for MagLS HRTF to degrade, yet enough residual energy for AA-MagLS HRTF to exploit in the optimization.
\color{black}

\color{black}
\section{Listening Experiment}

This section presents a listening experiment designed to subjectively evaluate the performance of the proposed ASM + AA-MagLS HRTF method against the benchmark ASM + MagLS HRTF, which represents state-of-the-art Ambisonics-based binaural reproduction.

\subsection{Setup}

The signals used for the listening test were generated using simulations as follows. A shoebox-shaped room was simulated using the  image method~\cite{Image-method-Allen1979} implemented in MATLAB \cite{MATLAB}. The room dimensions were $8 \times 6 \times 4$ meters (length × width × height), with a point source positioned at $(4, 3, 1.7)$ meters and a microphone array located at $(2.6, 4.4, 1.7)$ meters, about $2$ meters from the source. The microphone array was the same as described in Sec. \ref{section:Array Setup}, and illustrated in Figs\ref{fig:array_plot_2D} and \ref{fig:array_plot_3D}. The reverberation time was set to approximately $400$~ms, and the critical distance was estimated as $1.2$~meters.
A speech utterance taken from the TSP dataset \cite{kabal2002tsp}, sampled at $48,$kHz, was used as the audio signal and rendered in the simulated environment.

\subsection{Methodology}

\color{black}
To evaluate the proposed method in both ideal and practical scenarios, we compared several Ambisonics reproduction pipelines that differ in their encoding and processing configurations, array dependence, and HRTF preprocessing strategy. The selected methods represent (i) a theoretical upper bound under ideal conditions, (ii) a conventional low-order baseline, and (iii) array-dependent approaches reflecting real-world microphone geometries. This setup enables assessing how array-aware FOA encoding, such as ASM, perform in comparison to accurate FOA, and how encoding alternatives affect perceptual quality. The encoding methods under test are detailed in Sec.~\ref{section:Methodology}. Here is a description of the listening test signals:
\begin{itemize}
\item \textbf{HOA + HRTF}: Simulated high-order ($N=30$) Ambisonics convolved with a regular HRTF, as in Eq. (\ref{eq:p=hnmTanm}), provides the high quality reference.
\item \textbf{FOA + HRTF}: Low-order ($N=1$) Ambisonics convolved with a regular HRTF, as in Eq. (\ref{eq:p=hnmTanm}), sets the conventional low-order baseline used as anchor.
\item \textbf{FOA + MagLS HRTF}: Low-order Ambisonics combined with MagLS-optimized HRTFs, as described in Sec. \ref{sec:HRTFmagLS}, serves as the state-of-the-art benchmark using FOA signals.
\item \textbf{ASM + MagLS HRTF}: First order encoded Ambisonics combined with standard MagLS HRTF, as described in Sec. \ref{section: Ambisonics Encoding from Arbitrary Arrays}. This configuration represents the state-of-the-art encoded baseline.
\item \textbf{ASM + AA-MagLS HRTF}: The proposed method, as described in Sec. \ref{section:Proposed Method for binaural reproduction}.
\end{itemize}
\color{black}
The MUltiple Stimuli with Hidden Reference and Anchor (MUSHRA) test was used for evaluation under two conditions. In the first, the source was positioned approximately $45^\circ$ to the right, relative to the array front-looking direction, and without imposing head rotation on the HRTF. In the second condition, the array was rotated $60^\circ$ clockwise during capture, relative to the first condition. Then, a head rotation of $60^{\circ}$ was applied to the HRTF in the opposite direction, such that the source remained at the same direction relative to the listener head.

No head rotation was applied to the signals of the following methods: \textbf{HOA + HRTF}, \textbf{FOA + HRTF}, and \textbf{FOA + MagLS HRTF}, as the Ambisonics channels here are idealized and independent of the array. In total, seven distinct stimuli were generated.

Each listener compared the reference signal \textbf{HOA + HRTF} to the other signals, including the hidden reference. Note that \textbf{FOA + HRTF} was used as a lower anchor. For each scenario, two separate MUSHRA tests were conducted: one assessing spatial quality, including attributes such as localization accuracy, source direction perception, and externalization, \cite{lindau2014spatial} and the other assessing timbre, the spectral and temporal characteristics of the sound that contribute to its tonal color and texture. Each test included five stimuli (including a hidden reference and anchor) and required participants to rank similarity to the reference with respect to spatial quality or timbre. This resulted in four MUSHRA screens and a total of 20 ranked evaluations. A total of $12$ experienced participants, aged $25–40$ and with normal hearing, took part in the experiment.

\subsection{Results}

\begin{figure}[htbp]
    \centering
    \hspace*{-0.5cm}
    \includegraphics[width=0.55\textwidth]{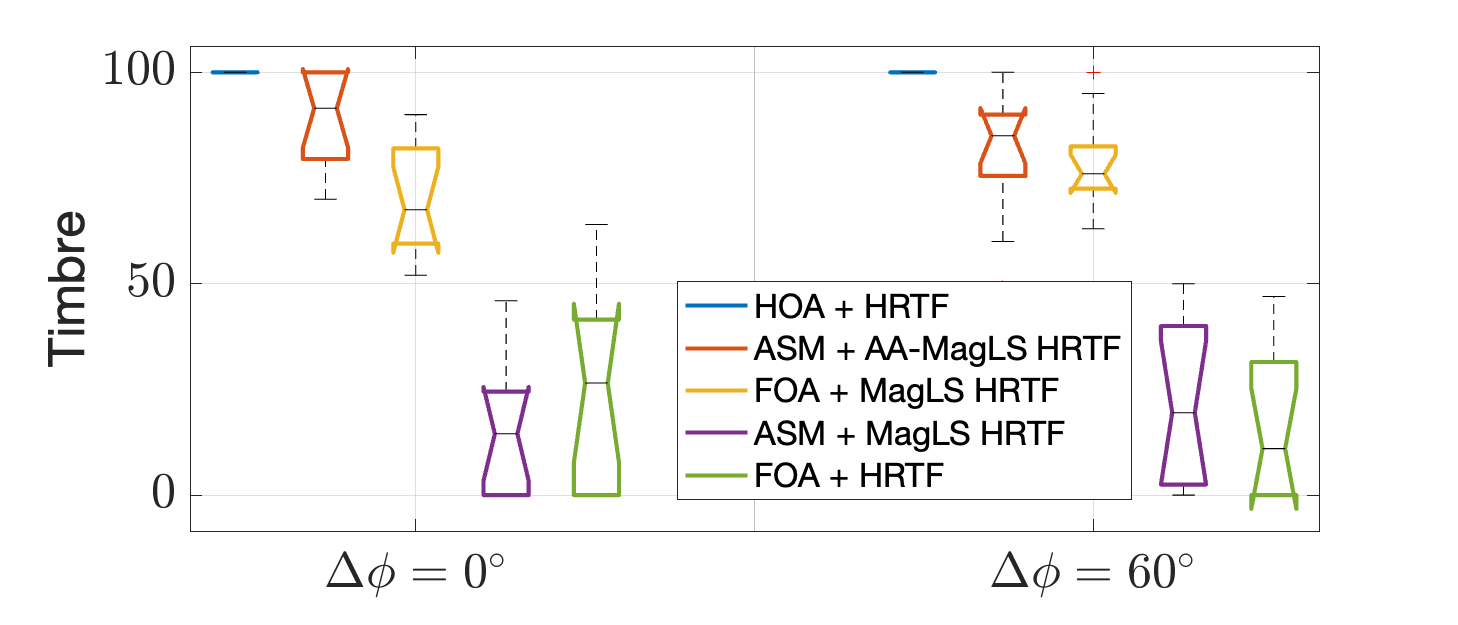}
    \hspace*{-0.5cm}
    \includegraphics[width=0.55\textwidth]{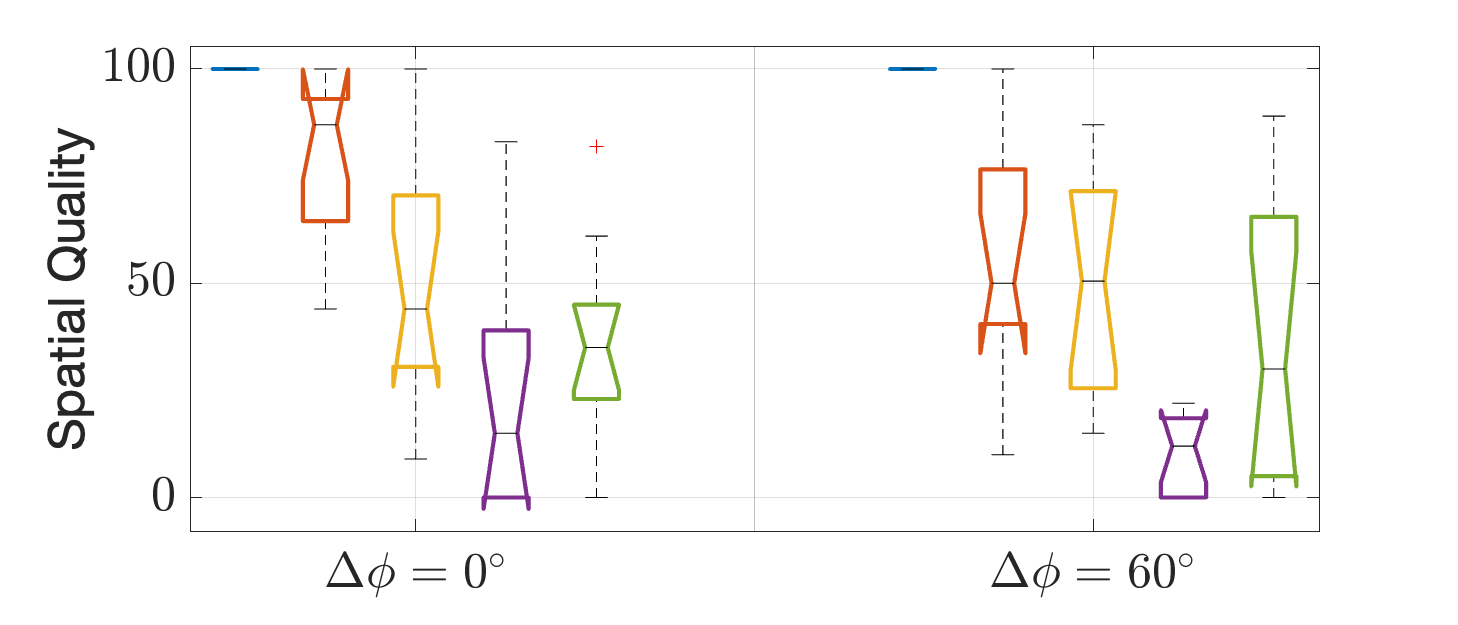}
        \caption{Box plots of the scores given by participants to each binaural reproduction method in the listening experiment, shown separately for \textit{timbre} (top) and \textit{spatial quality} (bottom) measures. Each subplot includes five reproduction methods: ASM + MagLS HRTF, ASM + AA-MagLS HRTF, FOA + HRTF, FOA + MagLS HRTF, and HOA + HRTF. The x-axis denotes the head rotation azimuth angle ($\Delta\phi = 0^\circ$ and $60^\circ$). For each box, the median score is marked by a horizontal red line; the 25th and 75th percentiles are represented by the lower and upper bounds of the box; whiskers indicate the minimum and maximum non-outlier values; and outliers are shown as red plus signs. Non-overlapping notches between boxes at the same angle indicate that the medians differ with 95\% confidence.}
    \label{fig:MUSHRA}
\end{figure}

The scores given by the participants to each test signal were analyzed using a repeated-measures ANOVA with two within-subject factors and their interaction: (1) the binaural reproduction method (ASM + AA-MagLS HRTF, ASM + MagLS HRTF, FOA + HRTF, HOA + HRTF, FOA + MagLS HRTF), and (2) the head orientation angle ($0^\circ$, $60^\circ$). The analysis was conducted separately for each perceptual measure: \textit{spatial quality} and \textit{timbre}, with $120$ observations per measure.

Mauchly’s test indicated sphericity violations for the method effect in the spatial quality measure ($p = .038$) and the method and head orientation interaction in the timbre measure ($p = .023$). In both cases, Greenhouse–Geisser corrections were applied, and all reported statistics reflect the corrected degrees of freedom. Sphericity was not violated for the method factor in the timbre measure ($p = .273$), and only marginally so for the interaction in the spatial measure ($p = .137$). No correction was required for the head orientation factor, which includes only two levels.

The analysis revealed a highly significant main effect of method, $F(2.12, \text{df}_\text{error}) = 36.29$, $p < .001$, indicating robust perceptual differences among the five reproduction methods. There was also a marginally significant interaction between method and head orientation, $F(2.65, \text{df}_\text{error}) = 3.01$, $p = .052$, suggesting that the effect of method was partially dependent on head orientation. The main effect of head orientation itself was not significant: $F(1, 11) = 3.29$, $p = .097$.

For the timbre measure, the main effect of method was again highly significant, $F(2.43, \text{df}_\text{error}) = 146.84$, $p < .001$, confirming large perceptual differences between methods. The main effect of head orientation was not significant: $F(1, 11) = 0.046$, $p = .834$. The method × head orientation interaction was also partially statistically significant: $F(2.24, \text{df}_\text{error}) = 2.95$, $p = .066$.

A box plot of results is visualized in Figure~\ref{fig:MUSHRA}, which displays participant scores for each reproduction method as a function of head orientation angle ($\Delta\phi = 0^\circ$ and $60^\circ$) and perceptual measure (\textit{timbre} and \textit{spatial quality}).

To further investigate the differences between binaural reproduction methods, Bonferroni-corrected pairwise comparisons were conducted, focusing on two key contrasts: (1) ASM + AA-MagLS HRTF vs.\ ASM + MagLS HRTF, and (2) ASM + AA-MagLS HRTF vs.\ FOA + MagLS HRTF. These comparisons were analyzed separately for the \textit{spatial quality} and \textit{timbre} measures.

In the \textit{timbre} measure, ASM + AA-MagLS HRTF showed a large and highly significant improvement over ASM + MagLS HRTF, with a mean difference of $+67.17$ ($p < .001$, 95\% CI: [47.78, 86.56]). When compared to FOA + MagLS HRTF, no statistically significant difference was observed ($p = .102$), with a smaller mean difference of $+11.17$ and a 95\% CI of [–1.45, 23.78], indicating comparable timbral performance. These results align with the analytical simulation in Sec.~\ref{section: Binaural Reproduction Error}, where the proposed ASM + AA-MagLS HRTF exhibits lower error than ASM + MagLS HRTF, as illustrated in Fig.~\ref{fig:HRTF Comparison with rotation}.

In the \textit{spatial quality} measure, ASM + AA-MagLS HRTF also outperformed ASM + MagLS HRTF, with a mean difference of $+49.59$ ($p < .001$, 95\% CI: [27.82, 71.37]), demonstrating statistically significant improvements in spatial reproduction. Similarly, its spatial performance was comparable to FOA + MagLS HRTF, showing no significant difference ($p = .119$), with a mean difference of $+18.05$ and a 95\% CI of [–2.93, 39.03]. It is interesting to note that the ILD and ITD analysis, as described in Sec.~\ref{section:ITD and ILD-Based Perceptual Evaluation Setup}. , did not show significant differences between ASM + AA-MagLS HRTF and ASM + MagLS HRTF. This discrepancy could be explained by the attenuation observed in the encoded Ambisonics signals (see Fig.~\ref{fig:ASM magnitude n=0,1}), which may suppress spatial information at high frequencies in ASM + MagLS HRTF. Such attenuation may not affect the ITD and ILD measures, as it may affect equally both ears.

These findings highlight that the proposed ASM + AA-MagLS HRTF method delivers substantial improvements over its baseline (ASM + MagLS HRTF), particularly in timbre, while also achieving notable gains in spatial quality. Moreover, its performance approaches that of FOA + MagLS HRTF in both timbre and spatial quality.

\section{Conclusions}

In this work, binaural reproduction methods for arbitrary microphone arrays were studied. The proposed ASM + AA-MagLS HRTF method jointly optimizes Ambisonics encoding and array-aware HRTF rendering, enabling more accurate binaural reproduction from wearable arrays. \color{black}The benefit of this array-aware formulation was shown to strongly depend on the array geometry and the resulting encoding errors. \color{black} Objective evaluations in simulated environments showed improved binaural accuracy compared to standard methods, particularly under head rotations and non-ideal microphone layouts.

These findings were supported by a controlled listening experiment, where participants consistently rated ASM + AA-MagLS HRTF higher in timbre compared to ASM + MagLS HRTF. Furthermore, ASM + AA-MagLS HRTF maintained robust perceptual performance across different head rotations.

Due to its low complexity and compatibility with standard Ambisonics pipelines, the proposed method may be especially useful for real-world applications such as augmented reality, virtual conferencing, and immersive media rendered from wearable recording devices.

\FloatBarrier
\newpage
\bibliographystyle{IEEEtran}
\bibliography{IEEEexample}

\end{document}